\documentclass[11pt,twoside]{./atmp}
\usepackage{epsfig}
\usepackage{amsfonts}
\usepackage{amscd}
\usepackage{graphics}
\usepackage{amsmath,amssymb}

\usepackage[all]{xy}
\usepackage{color}

\copyrightnotice{2005}{9}{0}{25} 

\def\rref#1{(\ref{#1})}
\def\be{\begin{equation}}
\def\beq{\begin{equation}}
\def\eeq{\end{equation}}
\def\ee{\end{equation}}
\def\bea{\begin{eqnarray}}
\def\eea{\end{eqnarray}}
\def\ba{\begin{array}}
\def\ea{\end{array}}

\def \um {\frac{1}{2}}

\def \a {\alpha}
\def \ga {\gamma}

\def\la{\lambda}
\def\ep{\epsilon}
\def\IZ{\mathbb{Z}}
\def\IR{\mathbb{R}}

\def\II{\mathbb{I}}

\begin{document}

\title[Constant connections]{Constant connections,\\ quantum holonomies 
and \\ the Goldman bracket}

\arxurl{math-ph/0412007}
\author[J.\ E.~Nelson and R.\ F.~Picken]{J.\ E.~Nelson${}^1$ and R.\ F.~Picken${}^2$}
\address{${}^1$ Dipartimento di Fisica Teorica, Universit\`a
degli Studi di Torino\\ and Istituto Nazionale di Fisica Nucleare, Sezione di
Torino\\  via Pietro Giuria 1, 10125 Torino, Italy\\ \vskip 0.3cm {nelson@to.infn.it}\\ 
\vskip 0.3cm
${}^2$ Departamento de Matem\'{a}tica and CAMGSD - \\
Centro de An\'{a}lise Matem\'{a}tica, Geometria e Sistemas Din\^{a}micos \\
 Instituto Superior T\'{e}cnico \\ 
Avenida Rovisco Pais, 1049-001 Lisboa, Portugal \\ \vskip 0.3cm{rpicken@math.ist.utl.pt}}

\vskip 0.3cm 

\begin{abstract}
In the context of $2+1$--dimensional quantum gravity with negative
cosmological constant and topology $\IR\!\times\!T^2$, constant matrix--valued
connections generate a $q$--deformed representation of the fundamental group,
and signed area phases relate the quantum matrices assigned to homotopic loops.
Some features of the resulting quantum geometry are explored, and as a
consequence a quantum version of the Goldman bracket is obtained. 
\vskip 0.3cm
\end{abstract}

\maketitle

\cutpage

\setcounter{page}{1}

\section{Introduction \label{sec1}}

Holonomy quantization in the context of
$2+1$--dimensional gravity with topology $\IR\!\times\!T^2$ and negative
cosmological constant $\Lambda$ has been discussed from a variety of points of
view, e.g. in terms of traced holonomies in \cite{NRZ,NR}, and in terms of the
holonomy matrices themselves \cite{NP1,NP2}. For further background see
\cite{car} and for a comparison with the second--order ADM approach see
\cite{CN}.

In previous articles \cite{NP1,NP2,NP3} we have studied $SL(2,\IR)$ holonomy 
matrices $U_1,\, U_2$ of the diagonal form 
\be
U_i = \left(\begin{array}{clcr}e^{{r_i}}& 0 \\0& e^{-{r_i}}\end{array}\right) 
\quad i=1,2
\label{hol}
\ee
considered as holonomies of the connection integrated along cycles $\ga_1,
\ga_2$ which generate the fundamental group of the torus, subject to the
relation 
\be
\gamma_1\cdot\gamma_2\cdot\gamma_1^{-1}\cdot\gamma_2^{-1}={\II}.
\label{fundga}
\ee
One of the reasons for studying matrices of the type \rref{hol} is that their 
gauge-invariant (normalized) traces $~ T_1^{\pm}=\um Tr U_1^{\pm}$ and 
$T_2^{\pm}=\um Tr U_2^{\pm}$ appear in $2+1$--dimensional gravity in various 
guises. In first--order holonomy quantization \cite{NRZ,NR} they satisfy the 
non--linear Poisson bracket algebra
\be
\{T_1^{\pm},T_2^{\pm}\}=\mp \frac{\sqrt {-\Lambda}}{4}(T_{12}^{\pm}-
 T_1^{\pm}T_2^{\pm}),
\label{b7}
\ee 
where in \rref{b7} the $\pm$ refer to the two copies (real and independent) of 
$\hbox{SL}(2,\IR)$ in
the isomorphism $\hbox{SO}(2,2)\cong\hbox{SL}(2,\IR)\otimes \hbox{SL}(2,\IR)$, and
$T_{12}^{\pm}=\um Tr (U_1^{\pm}U_2^{\pm})$ corresponds to the
loop $\gamma_1\cdot\gamma_2$. This loop has intersection number $-1$ with 
$\gamma_1$ and $+1$ with $\gamma_2$ (we assume the intersection number between 
$\ga_1$ and $\ga_2$ is $+1$.)

The traced holonomies of (\ref{b7}) can be represented classically as
\footnote{The parameters $r_{1,2}^{\pm}$ used here have been scaled by a factor of 
$\um$ with respect to \cite{NP1,CN}.} 
\beq
T_1^\pm = \cosh r_1^\pm , \quad T_2^\pm  = \cosh r_2^\pm , \quad
T_{12}^\pm = \cosh (r_1^\pm+r_2^\pm)  \label{cc6}
\eeq
where
$r_{1,2}^{\pm}$ 
are real, global, time-independent (but undetermined) 
parameters which, from (\ref{b7}), satisfy the Poisson brackets
\be
\{r_1^\pm,r_2^\pm\}=\mp \frac{\sqrt {-\Lambda}}{4}, 
\qquad \{r_{1,2}^+,r_{1,2}^-\}= 0.
\label{pb}
\ee

Alternatively, the traces $T_1^{\pm}$ and $T_2^{\pm}$ are the 
Wilson loops corresponding to $\ga_1$ and $\ga_2$ for ``shifted 
connections'' $\Gamma^{\pm}{}^{a}$ appropriate for $\Lambda < 0$ \cite{NM} and 
defined by\footnote{The third trace $T_{12}^{\pm}$ is determined from the 
identities $(T_1^{\pm})^2+(T_2^{\pm})^2+(T_{12}^{\pm})^2 - 2 
T_1^{\pm}T_2^{\pm}T_{12}^{\pm}=1$ which follow from the Mandelstam identity and
\rref{fundga} for the representations $U_1^{\pm},\, U_2^{\pm}$.}
\be
\Gamma^{\pm}{}^{a} = \omega^{a} \pm \sqrt{- \Lambda} ~e^{a},   
\label{gam}
\ee 
for suitable triads $e^{a}$ and spin connections 
$\omega^{a} = \frac{1}{2}\epsilon^{abc}\omega_{bc}, a,b,c=0,1,2$. The shifted 
connections \rref{gam} and therefore the traces $T_1^{\pm}, T_2^{\pm}$ can be 
calculated directly from the classical solutions \cite{CN}, in terms of the 
{\it same} parameters $r_1^\pm,r_2^\pm$. The explicit relationship is \cite{NM}
\be
(r_i^{\pm})^2= \Delta_i^{\pm}{}^a~\Delta_i^{\pm}{}^b~\eta_{ab},\quad i = 1,2, 
\quad \eta_{ab}= {\rm diag} (-1,+1,+1)
\label{rdel} 
\ee
where 
\be
\Delta_i^{\pm}{}^a = \int_{\ga_i} \Gamma^{\pm}{}^{(a)}.
\label{del}
\ee

We note that the Einstein--Hilbert action for (2+1)--dimensional gravity
\be
I_{\hbox{\scriptsize \it Ein}}
  = \int\- (d\omega^{ab}-{\omega^a}_d \wedge\omega^{db}
  +\frac{\Lambda}{3} e^a\wedge e^b)\wedge e^c\,\epsilon_{abc} ,
\qquad a,b,c=0,1,2.
\label{ein}\ee
can also be expressed (up to a total derivative) in terms of the connections 
\rref{gam} 
\be I_{\hbox{\scriptsize \it Ein}} = I^+ - I^-
\label{ein+}
\ee
where 
\be
I^{\pm}  = \frac{\a}{2}\int\- 
(d{\Gamma^{\pm}}_a - 
\frac{1}{3}{\Gamma^{\pm}}^b{\Gamma^{\pm}}^c\epsilon_{abc}){\Gamma^{\pm}}^a,
\qquad \Lambda = -  \frac{1}{\alpha^2}.
\label{eindel}
\ee

Quantization of the parameters satisfying \rref{pb} gives the commutators
\beq
[\hat r_1^\pm, \hat r_2^\pm] = \mp \frac{i\hbar \sqrt{-\Lambda}}{4}, \qquad 
[\hat r_{1,2}^+, \hat r_{1,2}^-] = 0.
\label{comm}
\eeq
It follows from \rref{comm}, and the identity
\be e^{\hat X} e^{\hat Y}= e^{\hat Y} e^{\hat X} e^{[ \hat X, \hat Y ]},
\label{bch}
\ee
valid when
$[ \hat X, \hat Y ]$ is a $c$--number, that the now {\it quantum} matrices 
$ {\hat U}_i^\pm$ \rref{hol} satisfy {\it by both matrix and operator 
multiplication}, the $q$--commutation relation
\be {{\hat U}_1}^{\pm} {{\hat U}_2}^{\pm}= 
q^{\pm 1} {{\hat U}_2}^{\pm} {{\hat U}_1}^{\pm}, 
\label{fund}
\ee
with
\be
q=\exp (- \frac {i \hbar \sqrt{-\Lambda}}{4})
\label{q}
\ee
i.e. a deformation of the classical equation which follows from 
\rref{fundga} and implies that the holonomies commute (see \cite{NP1}).

Quantum matrices in both the diagonal and upper--triangular sectors satisfying
the fundamental relation \rref{fund} have been studied in \cite{NP1,NP2}. We 
note that in order to make
the connection with $2+1$--dimensional gravity it is necessary to consider {\it
both} $\hbox{SL}(2,\IR)$ sectors, as, for example, in \rref{ein+}. The 
mathematical properties of just one
sector have been studied in \cite{NP3}. In fact one subsector of the classical
moduli space is when both holonomy matrices can be simultaneously conjugated
into diagonal form (conjugating both matrices by the same matrix $S\in
SL(2,\IR)$ ): 
\be 
U_i=\epsilon_i\left(\begin{array}{clcr}e^{{r_i}}& 0 \\0& e^{- {r_i}}
\end{array}\right) \quad \epsilon_i =\pm 1, \quad i=1,2. 
\ee

We will only consider the case $\epsilon_i=+1$, $i=1,2$,  i.e. the diagonal 
subsector containing the pair $(I,I)$, 

A convenient parametrization of flat connections for this diagonal sector 
was proposed in \cite{mik:pic}, namely by means of constant connections
\be
A= (r_1 dx + r_2 dy)
\left(\begin{array}{clcr} 1& 0 \\0& -1 
\end{array}\right)   
\label{conn}
\ee
where $x,y$ are coordinates, with period 1, on the torus 
$T^2= {\IR^2}_{(x,y)}/\IZ^2$, and $r_1$ 
and $r_2$ are global parameters. Now if $y$ is constant along
$\ga_1$ and  $x$ is  constant along $\ga_2$, the relation between the 
holonomy description \rref{hol} and the constant connection description is simply
\be
U_i = \exp \int_{\ga_i} A = \left(\begin{array}{clcr}e^{{r_i}}& 0 \\0& e^{-
{r_i}}\end{array}\right) \quad i=1,2.
\label{hol1}
\ee
In this diagonal sector the constant connection choice still allows some 
residual gauge freedom, since 
\bea
A' & = & 
\left(\begin{array}{clcr} 0& -1 \\1& 0 
\end{array}\right)^{-1} (d+A)
\left(\begin{array}{clcr} 0& -1 \\1& 0 
\end{array}\right) \nonumber \\
 & = & - (r_1 dx + r_2 dy)
\left(\begin{array}{clcr} 1& 0 \\0& -1 
\end{array}\right) = - A
\label{resgf}  
\eea
is gauge equivalent to $A$. So we may consider this diagonal part of the 
moduli space as parametrized by $(r_1,r_2)$, to be identified with 
$( - r_1, - r_2)$. The resulting conical structure is entirely in agreement with 
the holonomy description of \cite{NP3}.

By taking the $\hat r_i^\pm, i = 1,2$ to satisfy \rref{comm} the corresponding 
{\it quantum} holonomies ${\hat U}_i^\pm$ defined by \rref{hol1} satisfy the 
fundamental relation \rref{fund}. For completeness we note that there are also 
upper triangular solutions of \rref{fund} discussed in \cite{NP1}. These are
obtained in a similar fashion by defining ${\hat U}_i^\pm$ as in \rref{hol1},
but using constant quantum connections $A^{\pm}$ of the form 
\be A^{\pm} = {\la_1}^{\pm} dx + {\la_2}^{\pm} dy\label{j1}
\ee 
where the ${\la_i}^{\pm}$ are $2 \times 2$ matrices 
\be {\la_i}^{\pm} = \left(\begin{array}{clcr}{r_i}^{\pm}&\beta\\0& - {r_i}^{\pm}
\end{array}\right), \quad i = 1,2\label{j2}
\ee
whose internal entries satisfy ${r_i}^{\pm}\beta  + \beta {r_i}^{\pm}
= 0$
as well as \rref{comm}.

In this paper we investigate constant matrix--valued connections which
generalize the shifted connections \rref{gam}, and apply them to a much larger
class of loops. This leads to a definition of a $q$--deformed representation of
the fundamental group where signed area phases relate the quantum matrices
assigned to homotopic loops. The resulting structure is quantized and we
explore some features of the quantum geometry which arises. As a consequence we
obtain a quantization of the Goldman bracket \cite{gol}. The plan of the paper
is as follows.  In Section \ref{sec3} we assign quantum matrices to a large
class of loops represented by piecewise linear (PL) paths between integer
points in $\IR^2$, using a representation of $T^2$ as $\IR^2/\IZ^2$. We show
that the matrices for homotopic paths are related by a phase expressed in terms
of the signed area between the paths. In Section \ref{sec4} this observation
leads to the definition of a $q$-deformed surface group representation. In
Section \ref{sec5} we discuss the modular group action for the construction,
and in Section \ref{sec6} we use the PL representation of Section \ref{sec3} to
define intersections and reroutings, and quantize the Goldman bracket.

\noindent

\section{Quantum holonomies and area phases \label{sec3}}

\begin{figure}[hbtp]
\centering
\includegraphics[height=2cm]{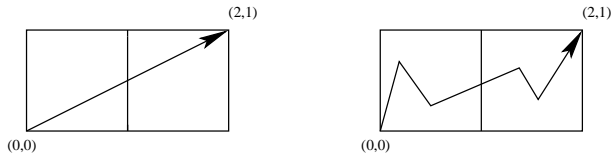}

\caption{Two PL paths in $\IR^2$ corresponding to homotopic loops}
\label{plexp}
\end{figure}

In this section we assign a quantum holonomy matrix to a large class of loops,
extending the assignments $\ga_1\mapsto U_1,\,\ga_2 \mapsto U_2$ by using the
quantum connection \rref{j1} in the diagonal case, i. e. with $\beta = 0$. We
only consider one sector, the $(+)$ sector of the classical moduli space.
Consider piecewise-linear (PL) paths on the plane $\IR^2$ going from the
origin $(0,0)$  to an integer point $(m,n), \, m,n\in \IZ$. Under the
identification $T^2=\IR^2/\IZ^2$, these paths give rise to closed loops on
$T^2$. The integers $m$ and $n$ are the winding numbers of the loop in the
$\ga_1$ and  $\ga_2$ directions respectively, and two loops on $T^2$ are
homotopic to each other if and only if the corresponding paths in $\IR^2$ end
at the same point $(m,n)$ - an example is shown in Figure \ref{plexp}. 

Suppose a PL path $p$ consists of $N$ straight segments $p_1, \dots, p_N$. Any
such segment $p_i$ may be translated to start at the origin and end at
$(a,b)\in \IR^2$  (here we use the fact that the connection $A$ is invariant
under spatial translations). Then we assign to each segment $p_i$ the quantum 
matrix

\be
\exp \int_{p_i} A = \exp \left((ar_1 + br_2)\sigma_3 \right) = 
\left(\begin{array}{cc} e^{ar_1+br_2}& 0 \\0& e^{-ar_1-br_2} 
\end{array}\right)\label{phi}   
\ee
where $\sigma_3 = \left(\begin{array}{clcr} 1& 0 \\0& -1 
\end{array}\right)$,  
and to the path $p$ the product matrix
\be
p\mapsto U_p := \prod _{i=1}^N \exp \int_{p_i} A.
\label{Up}
\ee

This assignment is obviously multiplicative under multiplication of
paths,  $(p,p')\mapsto p\circ p'$, which corresponds to translating $p'$ to
start at the endpoint of $p$ and concatenating.

\begin{figure}[hbtp]
\centering
\includegraphics[height=2cm]{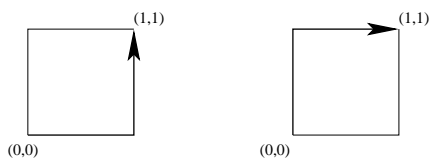}
\caption{The paths $p$ and $p'$ in the fundamental relation (\ref{fundsl2})}
\label{fundexp}
\end{figure}

Now denote the {\it straight} path from $(0,0)$ to $(m,n)$ by $(m,n)$; e.g. the first
path in Figure \ref{plexp} is $(2,1)$. In particular,  $U_1 = U_{(1,0)}$, $U_2
= U_{(0,1)}$, and these obey the fundamental relation
\be
U_1U_2= q U_2U_1,
\label{fundsl2}
\ee
with $q$ given by \rref{q}.
Geometrically we may regard this as a relation between $U_p$ and $U_{p'}$, 
where $p$ and $p'$ are the two paths shown in Figure \ref{fundexp}. 

\begin{figure}[hbtp]
\centering
\includegraphics[height=3cm]{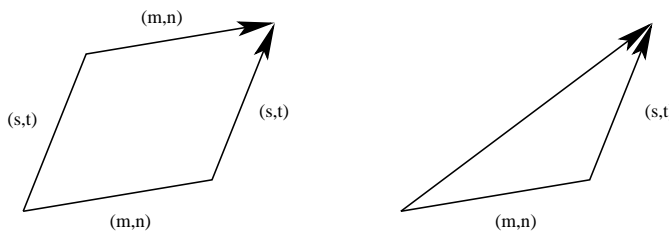}
\caption{Parallelogram and triangle for equations (\ref{uu1}) and (\ref{tri})}
\label{partri}
\end{figure}

Using (\ref{bch}) the fundamental relation (\ref{fundsl2}) may be 
generalized to arbitrary straight paths as follows:
\be
U_{(m,n)} U_{(s,t)} = q^{mt-ns} U_{(s,t)} U_{(m,n)},
\label{uu1}
\ee
where 
\be
U_{(m,n)} =
\left(\begin{array}{cc} e^{mr_1+nr_2}& 0 \\0& e^{-mr_1-nr_2} 
\end{array}\right).
\label{Umn}
\ee
Equation \rref{uu1} expresses the relation between the quantum matrices assigned 
to the two paths going from $(0,0)$ to $(m+s,n+t)$ in two different ways around 
the parallelogram generated by  $(m,n)$ and $(s,t)$, represented graphically in 
Figure \ref{partri}. It is also straightforward to show a triangle equation, 
also shown in Figure \ref{partri}
\be
U_{(m,n)} U_{(s,t)} = q^{(mt-ns)/2} U_{(m+s,n+t)},
\label{tri}
\ee
which can be derived from the identity
$$ e^{\hat X} e^{\hat Y}= e^{\hat X + \hat Y} e^{\frac{[\hat X, \hat Y ]}{2}},
$$
which follows from \rref{bch}. 

Note that in both cases the exponent of $q$ relating the two homotopic paths
is equal to the {\em signed area} between the path $p$ on the 
left-hand-side (l.h.s.)  and the path $p'$
on the right-hand-side (r.h.s.) 
 i.e. equal to the area between $p$ and $p'$, when the PL loop
consisting of $p$ followed by the inverse of $p'$ is oriented anticlockwise,
and equal to minus the area between $p$ and $p'$, when it is oriented
clockwise. The signed area for the parallelogram in Figure \ref{partri} is
given by 
$\det 
\left(\begin{array}{cc} m& s\\n& t 
\end{array}\right)
=mt-ns$
and for the triangle by $\um (mt-ns)$. 

\begin{figure}[hbtp]
\centering
\includegraphics[height=4cm]{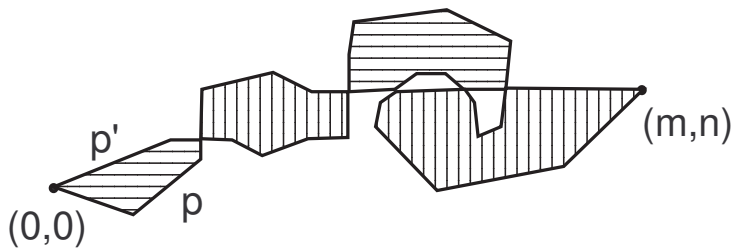}
\caption{ The $2$-chain $c(p,p')$}
\label{s_area_exp}
\end{figure}

We now generalize these observations to arbitrary non-self-intersecting PL
paths $p$ and $p'$ which connect $(0,0)$ to the same integer point $(m,n)$ in
$\IR^2$. These two paths may intersect each other several times, either
transversally, or when they coincide along a shared segment. Together they
bound a finite number of finite regions in the $xy$-plane. Now choose a
triangulation of a compact region of $\IR^2$ containing and compatible with the
paths $p,\,p'$, in the sense that each segment of the paths is made up of one
or more edges of the triangulation. We take all the triangles in the
triangulation to be positively oriented in the sense that their boundary is
oriented anticlockwise in $\IR^2$. Since $p$ and $p'$ are  homotopic, they are
homologous, and because $H_3$ of the plane is trivial, there is a unique
$2$-chain $c(p,p')$ such that $\partial c(p,p')=p-p'$. Let this chain be given
by 
\be
c(p,p') = \sum_{\alpha \in R}n_\alpha t_\alpha,
\label{2chain}
\ee
where $t_\alpha$ is a triangle of the triangulation indexed by $\alpha$ in the
index set $R$, and  $n_\alpha\ = \pm 1$ or $0$. Note that only triangles from
the finite  regions enclosed by $p$ and $p'$ can belong to the support of the 
$2$-chain, and that the coefficient of any two triangles in the same finite region
is the same. As an example, consider the two paths $p,\,p'$ in Figure
\ref{s_area_exp}. The $2$-chain  $c(p,p')$ is shown by the shaded regions. The
triangles inside the horizontally shaded  regions have $n_\alpha=+1$, and those
inside the vertically shaded regions  have $n_\alpha=-1$. The triangles inside
the white regions do not belong to the support of the  $2$-chain, i.e. they
have $n_\alpha =0$.

We now define the {\em signed area} between $p$ and $p'$ to be
\be
S(p,p') = \sum_{\alpha \in R} n_\alpha A(t_\alpha),
\label{spp}\ee
where 
$A(t_\alpha)$ is the area of the triangle $t_\alpha$. 
This is clearly independent of the
choice of  triangulation of $\IR^2$ compatible with $p,\,p'$, since the sum
of the areas of the  triangles inside each enclosed region is the area of
that region, whatever the  triangulation. 

We now have the main result of this section:  {\em for two paths $p,\,p'$ as
above the following relation holds}:  \[ U_p = q^{S(p,p')}U_{p'}. \]  This is
proven as follows.  Since $p$ and $p'$ are homologous we can construct a
sequence of  paths between them by collapsing or adding one triangle at a time.
Relation (\ref{tri}) for integer triangles generalizes to arbitrary  triangles,
and shows that the factor relating two successive paths in the sequence is
$q^{n_\alpha t_\alpha}$, where $t_\alpha$ is the triangle by which the two
successive paths differ. The overall  factor, after cancellations, is
determined by the chain $c(p,p')$.  


We note that the method of proof is reminiscent of Attal's combinatorial 
approach to gerbes and the four-colour problem \cite{att}. In principle an
alternative approach  would be to use the non-abelian Stokes theorem
\cite{nonabSt} and integrate the field  strength 
\[ 
F=dA + \um A \wedge A = [r_1, r_2] \II dx \wedge dy 
\] 
over the regions enclosed by $p$ and $p'$. Although the field strength is
non-vanishing, it commutes with the connection, which considerably simplifies
the non-abelian Stokes formula.

\section{$q$-deformed surface group representations \label{sec4}}

Classically there is a well-known correspondence between flat $G$-connections
on a manifold $M$ and group homomorphisms from $\pi_1(M)$ to the gauge group
$G$, up to equivalence (gauge equivalence for the connections and conjugation
by a fixed element of $G$ for the group homomorphisms). Since here $M$
is a surface, such homomorphisms are known as {\em surface group
representations}. Because of the constant connection gauge
choice, the gauge equivalence for the connections is virtually trivialized,
apart from the residual gauge freedom of equation (\ref{resgf}), and
the group $G$ is a diagonal subgroup of $SL(2,\IR)$.

Let $\Omega T^2$ denote the group of closed loops on $T^2$ identified with
the PL paths $p$ from $(0,0)$ to an integer point in $\IR^2$ introduced in the
previous section. Then a surface group representation corresponds to an
assignment $\phi$ to each $p \in \Omega T^2$ of an element
$\phi(p)\in G$, such that

\begin{itemize}
\item[1)] $p_1 \sim p_2 \Longrightarrow \phi(p_1)= \phi(p_2)$ 
(where $\sim$ denotes ``is homotopic to'') 
\item[2)] $\phi(p_1p_2)= \phi(p_1)\phi(p_2)$.
\end{itemize}
Then a constant classical connection of the form $A= (r_1 dx + r_2 dy)\sigma_3$
gives rise to an assignment $ \phi_{r_1,r_2}$, namely
\[
\phi_{r_1,r_2}(p) =\exp \int_p A = \exp \left((mr_1 + nr_2)
\sigma_3\right)
\]
for any path $p$ connecting $(0,0)$ to $(m,n)$ 
(these are all homotopic).

\begin{figure}[hbtp]
\centering
\includegraphics[height=4cm]{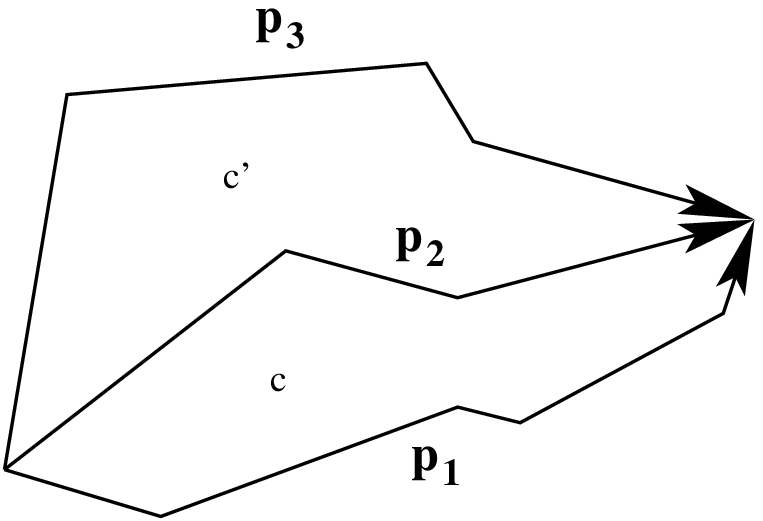}
\caption{Consistency of equations (\ref{qdsgeq1}) and (\ref{qdsgeq2})}
\label{qdsg1}
\end{figure}

\begin{figure}[!ht]
\centering
\includegraphics[height=4cm]{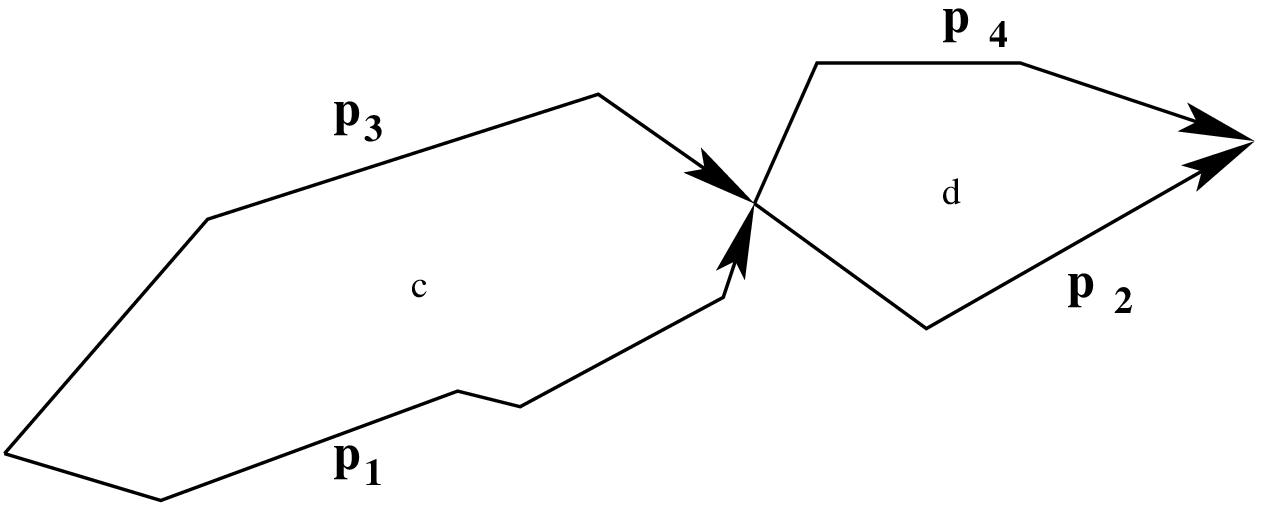}
\caption{Consistency of equations (\ref{qdsgeq3}) and (\ref{qdsgeq4})}
\label{qdsg2}
\end{figure}

Now we define a $q$-deformed version of the above appropriate to the constant
quantum connections  of Section 3. Let $\hat{G}$ denote
the quantum matrices generated multiplicatively by matrices of the form
\rref{phi}
 $$ \exp \left((ar_1 + br_2) \sigma_3 \right)$$
where $a,b\in \IR$. 
A {\em $q$-deformed surface group representation} is an assignment $\phi$
to each $p \in \Omega T^2$ of an element $\phi(p)\in \hat{G}$, such 
that
\begin{itemize}
\item[1)] $p_1 \stackrel{c}{\sim} p_2 \Longrightarrow \phi(p_1) 
= q^{S(p_1,p_2)}\phi(p_2)$ where $\stackrel{c}{\sim} $ denotes ``is 
homotopic to, with the paths $p_1,\,p_2$ in $\IR^2$ differing by the 
$2$-chain $c$ of equation \rref{2chain}'' 
\item[2)] $\phi(p_1p_2)= \phi(p_1)\phi(p_2)$.
\end{itemize}
This is well-defined. Firstly, if $p_1 \stackrel{c}{\sim} p_2$ and 
$p_2 \stackrel{c'}{\sim} p_3$, we have
\be
\left. \begin{array}{lcr} \phi(p_1) & = & q^{S(p_1,p_2)}\phi(p_2) \\
\phi(p_2) & = & q^{S(p_2,p_3)}\phi(p_3)
\end{array} \right\} \Longrightarrow \phi(p_1)  =  
q^{S(p_1,p_2)+S(p_2,p_3)}\phi(p_3)
\label{qdsgeq1}
\ee
but also
\be
\phi(p_1)  =  q^{S(p_1,p_3)} \phi(p_3).
\label{qdsgeq2}
\ee
Now since $\partial c =p_1-p_3, \, \partial c' = p_2 -p_3$ and therefore
$\partial (c+c') =p_1-p_3$, it follows that 
$p_1 \stackrel{c+c'}{\sim} p_3$ (see Figure \ref{qdsg1}). 
Then equations (\ref{qdsgeq1}) and (\ref{qdsgeq2}) are 
consistent since, writing $c= \sum_{\alpha \in R} n_\alpha t_\alpha$, 
$c'= \sum_{\alpha \in R}n'_\alpha t_\alpha$, and therefore 
$c + c'= \sum_{\alpha \in R} (n_\alpha + n'_\alpha)t_\alpha$, 
we have, from \rref{spp}, 
\begin{eqnarray}
S(p_1,p_3) & = & \sum_{\alpha \in R}(n_\alpha + n'_\alpha)A (t_\alpha)\nonumber \\
& = &  \sum_{\alpha \in R} n_\alpha A(t_\alpha) +
\sum_{\alpha \in R} n'_\alpha A(t_\alpha)\nonumber\\
& = & S(p_1,p_2) + S(p_2,p_3).
\end{eqnarray}
Secondly, let $p_1 \stackrel{c}{\sim} p_3$ and 
$p_2 \stackrel{c'}{\sim} p_4$, i.e. $\partial c= p_1-p_3$, 
$\partial c' =p_2-p_4$. Let $d$ be the $2$-chain obtained from 
$c'$ by shifting the triangles by the vector $(m,n)$, the common endpoint of 
$p_1$ and $p_3$ (see Figure \ref{qdsg2}). Then 
\[
\partial(c+ d) = p_1p_2 - p_3p_4
\]
 
\noindent and
\begin{eqnarray}
\phi(p_1p_2) & = & \phi(p_1) \phi(p_2) \nonumber \\
& = & q^{S(p_1,p_3)} q^{S(p_2,p_4)} \phi(p_3) \phi(p_4) \nonumber \\
& = & q^{S(p_1,p_3) + S(p_2,p_4)} \phi(p_3 p_4)
\label{qdsgeq3}
\end{eqnarray}
but also 
\be
\phi(p_1 p_2) = q^{S(p_1p_2,p_3p_4)} \phi(p_3 p_4).
\label{qdsgeq4}
\ee
Equations (\ref{qdsgeq3}) and (\ref{qdsgeq4}) are consistent, since writing 
\[
c= \sum_{\alpha \in R} n_\alpha t_\alpha
\quad 
c'= \sum_{\alpha \in R} n'_\alpha t_\alpha
\quad
d= \sum_{\alpha \in R} \tilde{n}'_\alpha t_\alpha
\]
it follows that 
\[
S(p_1,p_3) + S(p_2,p_4) = \sum_{\alpha \in R} ( n_\alpha + {n}'_\alpha) A(t_\alpha)
\]
and
\begin{eqnarray*}
S(p_1p_2,p_3p_4) & = & \sum_{\alpha \in R} (n_\alpha + 
\tilde{n}'_\alpha) A(t_\alpha) \\
& = & \sum_{\alpha \in R} (n_\alpha + {n}'_\alpha) A(t_\alpha) \\
& = & S(p_1p_2,p_3p_4),
\end{eqnarray*}
since the area is clearly unaffected by the translation.

By analogy with the classical case, a constant quantum connection
\[
A=(r_1 dx + r_2 dy)\sigma_3
\]
gives rise to a $q$-deformed surface group representation $\phi_{r_1,r_2}$ by:
\[
\phi_{r_1,r_2}(p) = U_p
\]
where $U_p$ was defined in 
\rref{Up}. From the results of Section 3 (the  
multiplicativity of the assignment $p\mapsto U_p$, and the area phases
relating the matrices for homotopic paths) it is clear that
 $\phi_{r_1,r_2}$ satisfies 
the conditions for being a $q$-deformed surface group representation. 

\section{The Modular Group\label{sec5}}

Consider the piecewise linear (PL) paths of Section \ref{sec3}. Since $SL(2,\IZ)$ 
acts on $\IR^2$ by multiplication, fixes the origin and preserves the lattice of
points $\IZ^2 \in \IR^2$, it maps straight segments to straight segments
and PL paths to PL paths. This defines an action of 
$SL(2,\IZ)$ on the collection of matrices $U_p$ of equation \rref{Up} by
\be 
M.U_{p} = U_{M.p}.
\ee
For straight paths $(m,n)$ - for example
those appearing in equation \rref{fundsl2} - we can make contact with earlier work 
\cite{NP1,NP2} by setting $U_1 = U_{(1,0)}$, $U_2 = U_{(0,1)}$ and checking the
action of the standard generators of $SL(2,\IZ)$ 
\be
T = \left(\begin{array}{clcr}1&0\\1&1\end{array}\right), \qquad
S = \left(\begin{array}{clcr}0&-1\\1&0\end{array}\right) 
\label{ts}
\ee
on $U_1$ and $U_2$, namely 
\bea
T.U_1 &=& T.U_{(1,0)} =  U_{(1,1)}= q^{-1/2} U_1U_2\nonumber \\
T.U_2 &=& T.U_{(0,1)} =  U_{(0,1)}= U_2 \nonumber \\
S.U_1 &=& S.U_{(1,0)} =  U_{(0,1)}= U_2\nonumber \\
S.U_2 &=& S.U_{(0,1)} =  U_{(-1,0)}= U_1^{-1}
\label{stact}
\eea
which preserve the fundamental relation \rref{fund}. In the first of
\rref{stact} (the equation for $T.U_1$) the triangle relation \rref{tri} was
used. The first two of \rref{stact} correspond to the more familiar {\it
classical} transformations
\be
U_1 \stackrel{T}{\to} U_1U_2, \qquad U_2 \stackrel{T}{\to} U_2,
\ee
whereas the last two correspond to
\be
U_1 \stackrel{S}{\to} U_2, \qquad U_2 \stackrel{T}{\to} U_1^{-1},
\ee
which differ from \rref{stact} by a phase, but still preserve \rref{fund}. 
Note that in both cases $S^2 = -\II, S^4 = \II, (ST)^3 = \II$, both
algebraically and geometrically. For example, algebraically 
$T.U_1^{-1}=q^{1/2} U_2^{-1}U_1^{-1}$ whereas geometrically
$T.U_1^{-1}=T.U_{(-1,0)} = U_{(-1,-1)}$. These are equal when the triangle 
relation \rref{tri} is applied.

We can also define a dual action of $SL(2,\IZ)$ on the $q$-deformed
 surface group representations of Section \ref{sec4} for generic paths $p$ by
\be
(M.\phi)(p) = \phi(M^{-1}.p). 
\ee
\noindent This is a left action in the usual way since
\bea 
(M_1.(M_2.\phi))(p) &=& (M_2.\phi)(M_1^{-1}.p)\nonumber \\
&=& \phi(M_2^{-1}.(M_1^{-1}.p))\nonumber \\
&=& \phi(M_2^{-1}M_1^{-1}).p)) \nonumber \\  
&=& \phi((M_1M_2)^{-1}.p))\nonumber \\
&=& (M_1M_2.\phi)(p).  
\eea
It is also well defined since, for $ p_1 \stackrel{c}{\sim} p_2$, 
\bea
(M.\phi)(p_1)& = & \phi(M^{-1}.p_1) \nonumber  \\
 &= &   q^{S(M^{-1}.p_1,M^{-1}.p_2)}\phi(M^{-1}.p_2) \nonumber \\
& = & q^{S(p_1,p_2)} (M.\phi)(p_2)
\eea
since $M\in SL(2,\IZ)$ has determinant $1$ and thus preserves areas. Moreover,
\bea
(M.\phi)(p_1 p_2) &=& \phi(M^{-1}.(p_1 p_2))  \nonumber \\
& = & \phi((M^{-1}.p_1)(M^{-1}.p_2)) \nonumber \\
 &=& \phi(M^{-1}.p_1)\phi(M^{-1}.p_2)  \nonumber \\
&=& (M.\phi)(p_1)(M.\phi)(p_2).  
\eea

\section{The Goldman bracket \label{sec6}} There is a classical bracket due to
Goldman \cite{gol} for functions $T(\ga)={\rm tr}\, U_\ga$ defined on
homotopy classes of loops $\ga$, which for $U_\ga \in SL(2,\IR)$ is (see
\cite{gol} Thm. 3.14, 3.15 and Remark (2), p. 284):
\be
\{T(\ga_1), T(\ga_2)\} = \sum_{S \in \ga_1 \sharp \ga_2}
\epsilon(\ga_1,\ga_2,S)(T(\ga_1S\ga_2) -
T(\ga_1S\ga_2^{-1})).
\label{gold}
\ee

Here $\ga_1 \sharp \ga_2$ denotes the set of (transversal) intersection points
of $\ga_1$ and $\ga_2$ and $\epsilon(\ga_1,\ga_2,S)$ is the intersection index
for the intersection point $S$.  $\ga_1S\ga_2$ and $\ga_1S\ga_2^{-1}$ denote
loops which are  {\it rerouted} at the intersection point $S$. In the
following we explain and use  these concepts, and show how formula
\rref{gold} appears in the context of the PL paths introduced in Section
\ref{sec3}. Equation \rref{gold} may be quantized in two different but
equivalent forms using the concept of area phases for homotopic paths  which
was developed in Sections \ref{sec3} and \ref{sec4}.

\subsection{The Fundamental Reduction\label{fundred}} In order to study
intersections of ``straight'' loops, represented in $\IR^2$ by straight paths
between $(0,0)$ and integer points $(m,n)$, it is useful to consider their
reduction to a fundamental domain of $\IR^2$, namely the square with vertices
$(0,0), (1,0), (1,1), (0,1)$.

\begin{figure}[hbtp]
\centering
\includegraphics[height=2cm]{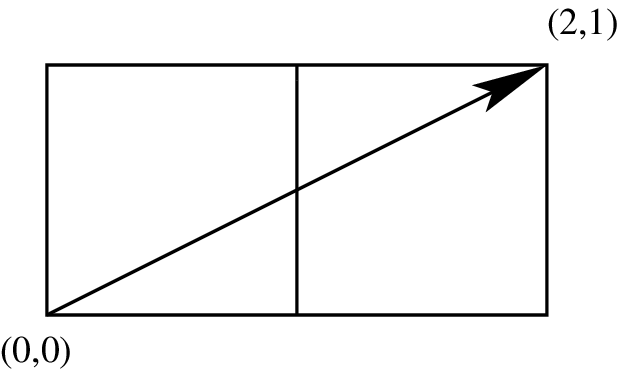}
\hspace{2cm}
\includegraphics[height=2cm]{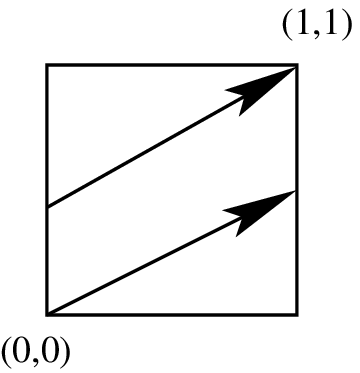}
\caption{The path $(2,1)$ and its fundamental reduction}
\label{21}
\end{figure}

We give some examples of fundamental reduction. Figure \ref{21} shows a path
in the first quadrant, namely the path $(2,1)$, and its reduction to the
fundamental domain

\begin{figure}[hbtp]
\centering
\includegraphics[width=2cm]{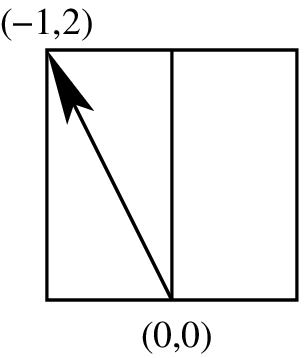}
\hspace{2cm}
\includegraphics[width=2cm]{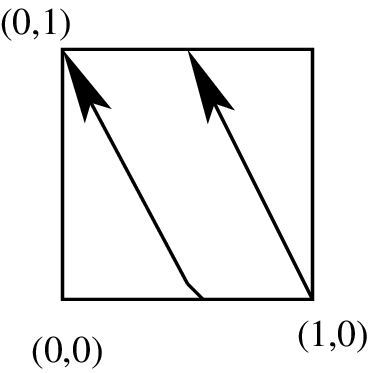}
\caption{The path $(-1,2)$ and its fundamental reduction}
\label{-12}
\end{figure}

\begin{figure}[ht]
\centering
\includegraphics[width=2cm]{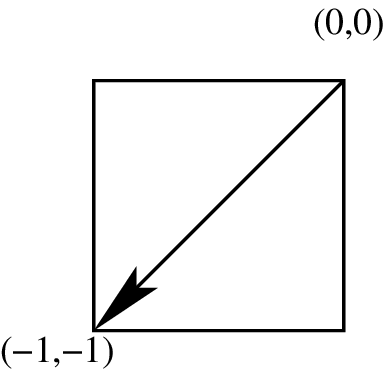}
\hspace{2cm}
\includegraphics[width=2cm]{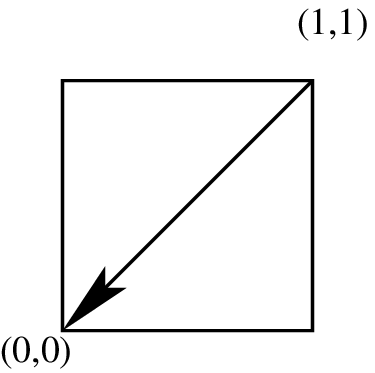}
\caption{The path $(-1,-1)$ and its fundamental reduction}
\label{-1-1}
\end{figure}

\begin{figure}[!ht]
\centering
\includegraphics[width=3.5cm]{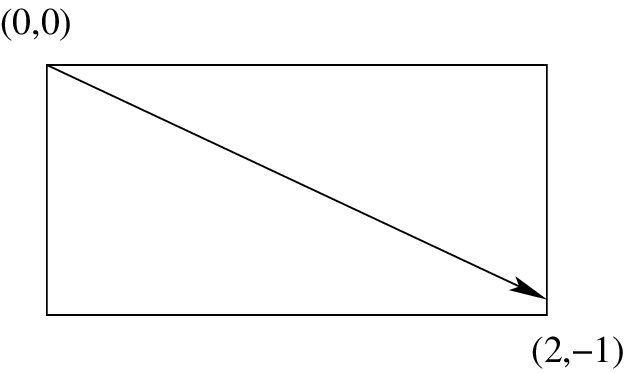}
\hspace{2cm}
\includegraphics[width=2cm]{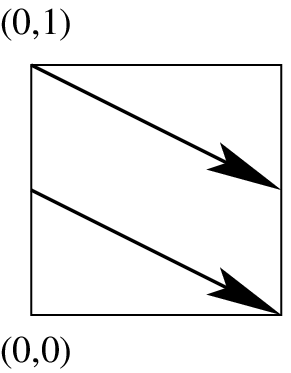}
\caption{The path $(2,-1)$ and its fundamental reduction}
\label{2-1}
\end{figure}

\noindent whereas in other quadrants fundamentally reduced paths start at 
other vertices (not $(0,0)$). For example in the second quadrant the path 
$(-1,2)$ will (in the fundamental domain) start at $(1,0)$ and end at $(0,1)$, 
as shown in Figure \ref{-12}.  
\noindent Examples of the reduction of paths from the third quadrant (the path
$(-1,-1)$, starting at $(1,1)$ and ending at $(0,0)$) and the fourth quadrant
(the path $(2,-1)$, starting at $(0,1)$ and ending at $(1,0)$) are shown in
Figures \ref{-1-1} and \ref{2-1} respectively.

\begin{figure}[!ht]
\centering
\includegraphics[width=2cm, height=3cm]{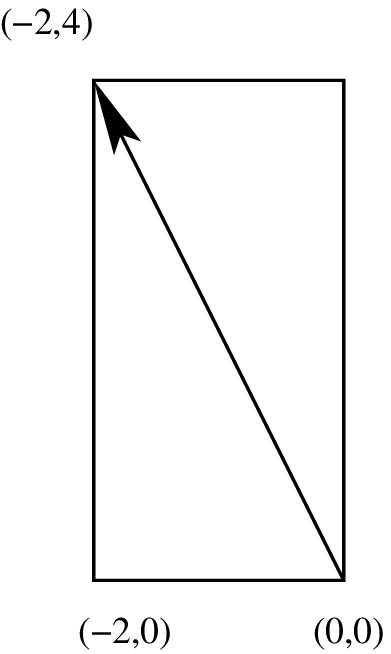}
\hspace{2cm}
\includegraphics[width=2cm]{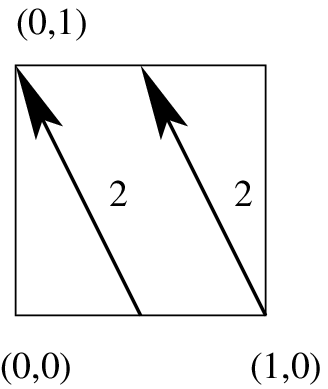}
\caption{The path $(-2,4)$ and its fundamental reduction}
\label{-24}
\end{figure}

When the path $(m,n)$ is a multiple of another integer path, we say it is
{\it reducible}. Otherwise it is irreducible. Figure \ref{-24}
shows how this multiplicity is indicated in the fundamental reduction.

\subsection{Intersections\label{ints}}

It is clear from Section \ref{fundred} that two paths intersect at any points
where their fundamental reductions intersect. We may only consider transversal
intersections, namely when their respective tangent vectors are not collinear.
For paths of multiplicity $1$ intersecting at a point, we say that their
intersection number at that point is $+1$ if the angle from the first tangent
vector to the second is between $0$ and $180$ degrees, and $-1$ if between
$180$ and $360$ degrees. For paths of multiplicity greater than $1$, the
intersection number is multiplied by the multiplicities of the paths involved.
We denote the intersection number between two paths $p_1$ and $p_2$ at a point
$P$ (or $P, Q, R$ if more than one) by  $\epsilon(p_1,p_2,P)$. The total
intersection number for two paths is the sum of  the intersection numbers for
all the intersection points, denoted  $\epsilon(p_1,p_2)$. We list some simple
(and not so simple) examples (in the fundamental domain) of single and multiple
intersections.  
\begin{enumerate}
\item If $p_1= (1,0)$ and $p_2=(0,1)$ then there is a single 
intersection at the point $(0,0)$ with $\ep=+1$.
\hfill\includegraphics[width=2cm]{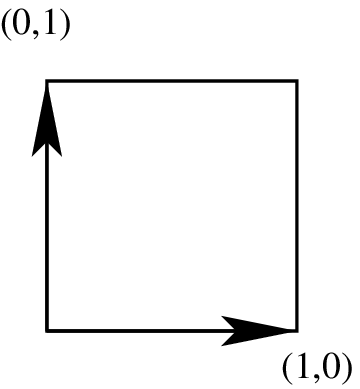}

\item If $p_1= (2,1)$ and $p_2=(0,1)$ then there are two 
intersections, one at the point $P=(0,0)$, and one at the point
$Q=(0,\um)$, each with $\ep=+1$, so the total intersection number is 
$\ep=+2$
\hfill\includegraphics[width=2cm]{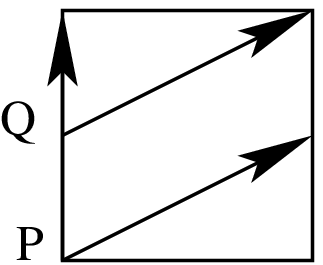}
 
\item If $p_1= (1,2)$ and $p_2=(2,1)$ then there are three  intersections,
one at the point $P=(0,0)$, and two others at the points
$Q=(\frac{2}{3},\frac{1}{3})$ and $R(\frac{1}{3},\frac{2}{3})$ (see figure),
each with $\ep=-1$, so the total intersection number is $\ep=-3$ (the point 
$S=(1,1)$ does not contribute since it coincides with the point $P$)
\hfill\includegraphics[width=2cm]{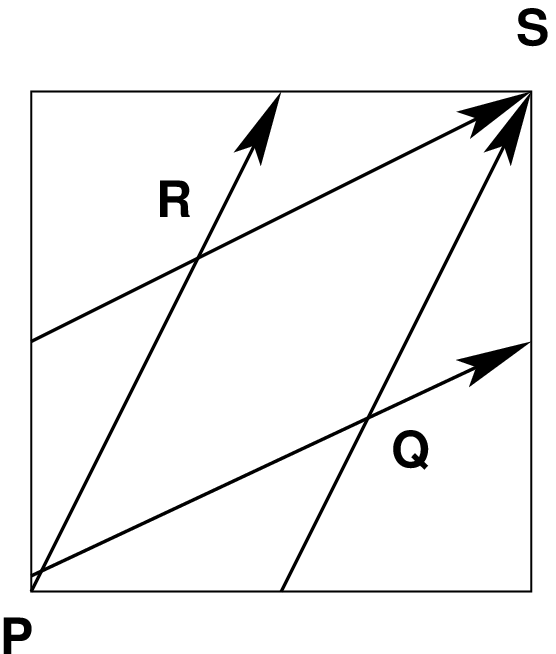}

\item If $p_1= (1,1)$ and $p_2=(-1,2)$ then there are three  intersections,
one at the point $P=(0,0)$, and two others at the points 
$Q=(\frac{2}{3},\frac{2}{3})$ and $R=(\frac{1}{3},\frac{1}{3})$ (see figure),
each with $\ep=+1$, so the total intersection number is $\ep=+3$
\hfill\includegraphics[width=2cm]{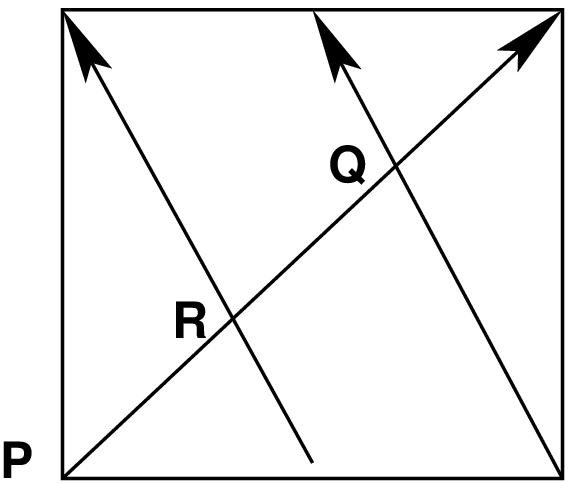}

\end{enumerate}

We remark that 
\begin{enumerate}
\item all intersections between a given pair of straight paths have the same
sign, since in this representation their tangent vectors have constant
direction along the loops. 

\item the total intersection number between $p_1=(m,n)$ and $p_2=(s,t)$ 
is given by the determinant 
\be
\ep(p_1,p_2) =  \left |\ba{clcr}m&n\\s&t\ea
\right | = mt - ns
\label{detint}
\ee
This is shown by using the fact that the total intersection number is invariant 
under deformation, i.e. homotopy
\begin{eqnarray}
\ep((m,n),(s,t)) & = & \ep((m,0)+(0,n), (s,0)+(0,t)) \nonumber \\
& = & \ep((m,0),(0,t)) + \ep((0,n),(s,0)) \nonumber \\
& = & mt-ns.
\end{eqnarray}
Relation \rref{detint} is easily checked for the above examples.

\end{enumerate}

\subsection{Reroutings\label{rerout}}

Consider two straight paths $p_1$ and $p_2$ and let $P$ be a point at
which they intersect.  The positive and negative reroutings are the paths
denoted by $p_1Pp_2$ and  $p_1Pp_2^{-1}$ respectively, where
$p_2^{-1}= (-s,-t)$ if  $p_2=(s,t)$. These reroutings are defined as
follows: starting at the basepoint follow $p_ 1$ to $P$, continue on $p_2$
(or $p_2^{-1}$) back to $P$, then finish along $p_ 1$.  Note that, in
accordance with the above rule, if the intersection point $P$ is the basepoint
itself, the reroutings $p_1Pp_2$ and  $p_1Pp_2^{-1}$ start by following
$p_2$ (or $p_2^{-1}$) from the basepoint back to the basepoint, and then
follow $p_1$ from the basepoint back to the basepoint. Here we show the
reroutings ($p_1Pp_2$ and $p_1Pp_2^{-1}$ respectively, and at the
various intersection points $P, Q, R$ if more than one) for the four examples
of Section \ref{ints}, using the non--reduced paths which are more convenient
here. 

\begin{enumerate}

\item $P=(0,0), p_1 = (1,0), p_2 = (0,1)$ 
\hfill
\includegraphics[width=2cm]{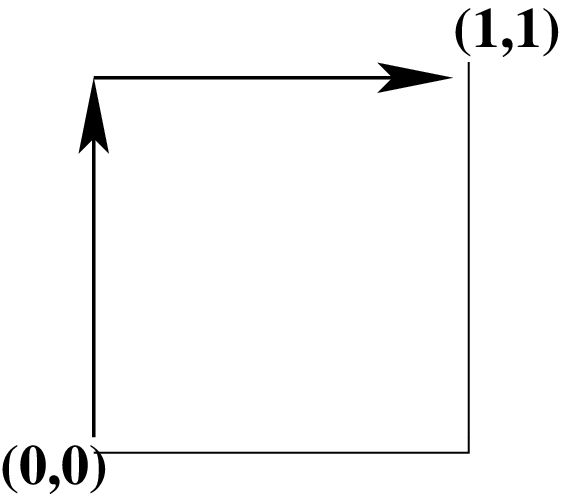}
\hspace{0.5cm}
\includegraphics[width=2.2cm]{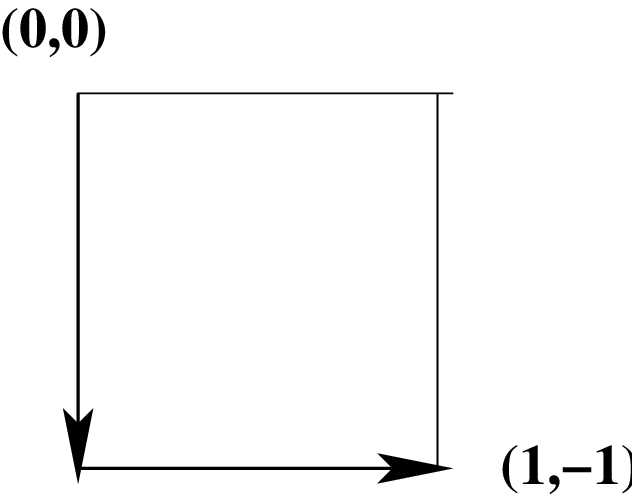}

\item (a)~$P=(0,0), p_1 = (2,1), p_2 = (0,1)$
\hfill
\includegraphics[width=2cm]{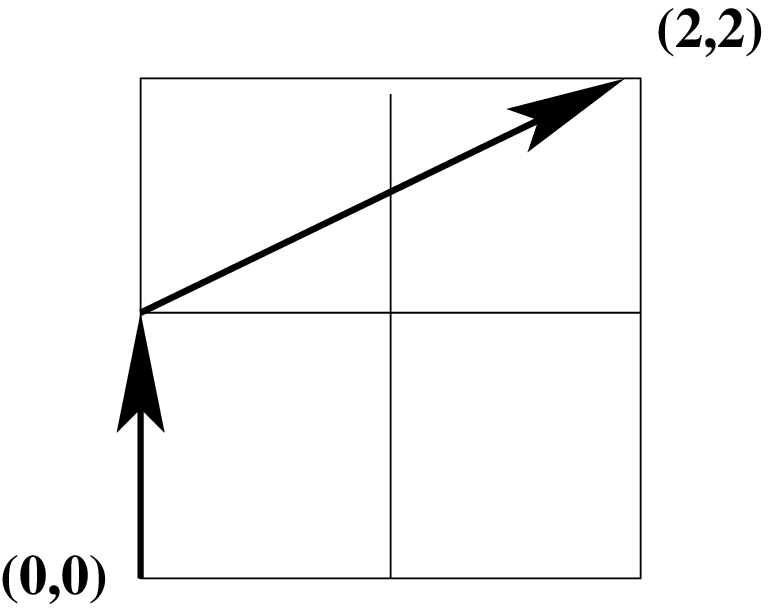}
\hspace{0.5cm}
\includegraphics[width=2cm]{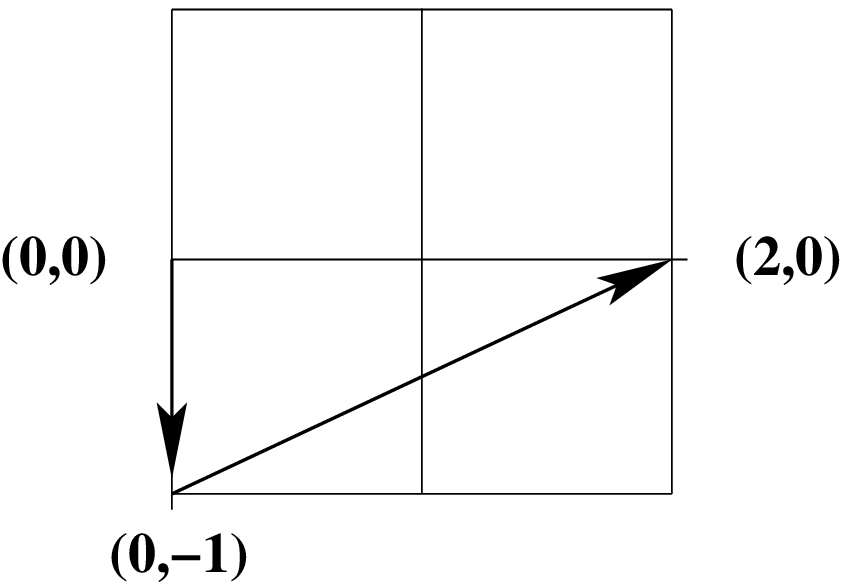}
\vspace{.5cm}
\noindent (b)~$Q=(0,\um), p_1 = (2,1), p_2 = (0,1)$
\hfill
\includegraphics[width=2cm]{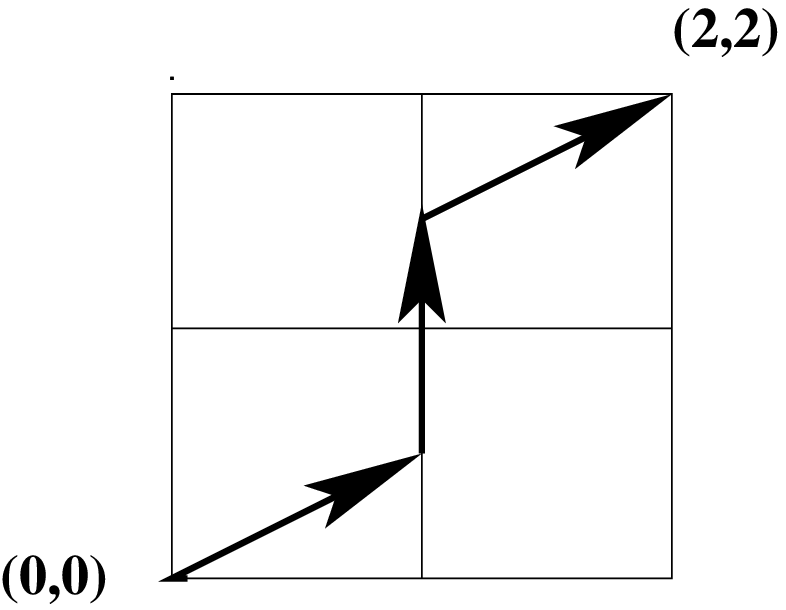}
\hspace{0.5cm}
\includegraphics[width=2cm]{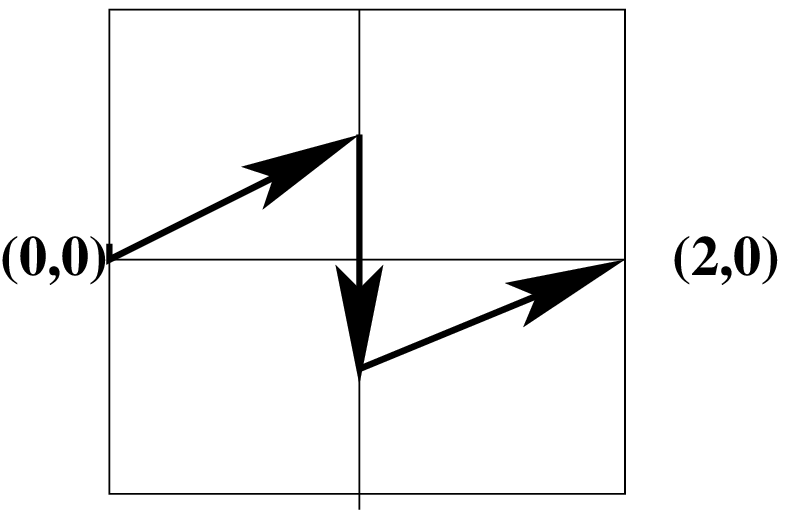}

\item (a)~$P=(0,0), p_1 = (1,2), p_2 = (2,1)$
\hfill
\includegraphics[width=2.5cm]{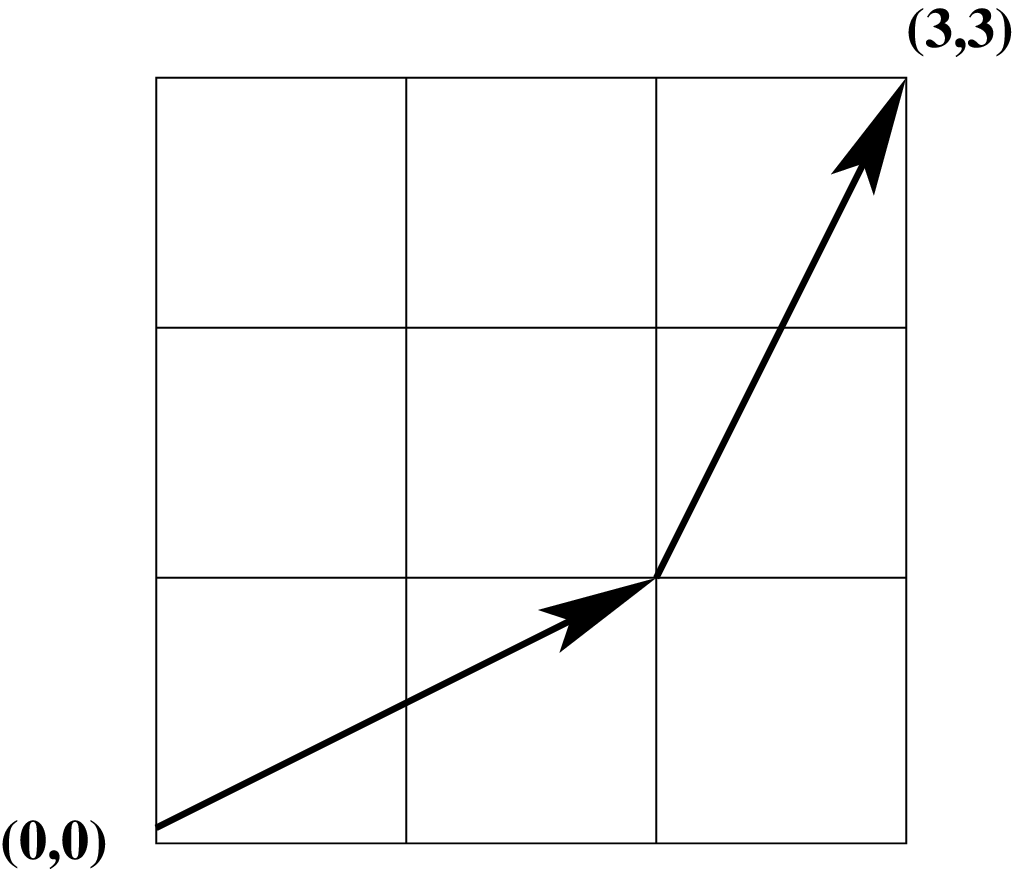}
\hspace{0.5cm}
\includegraphics[width=2cm]{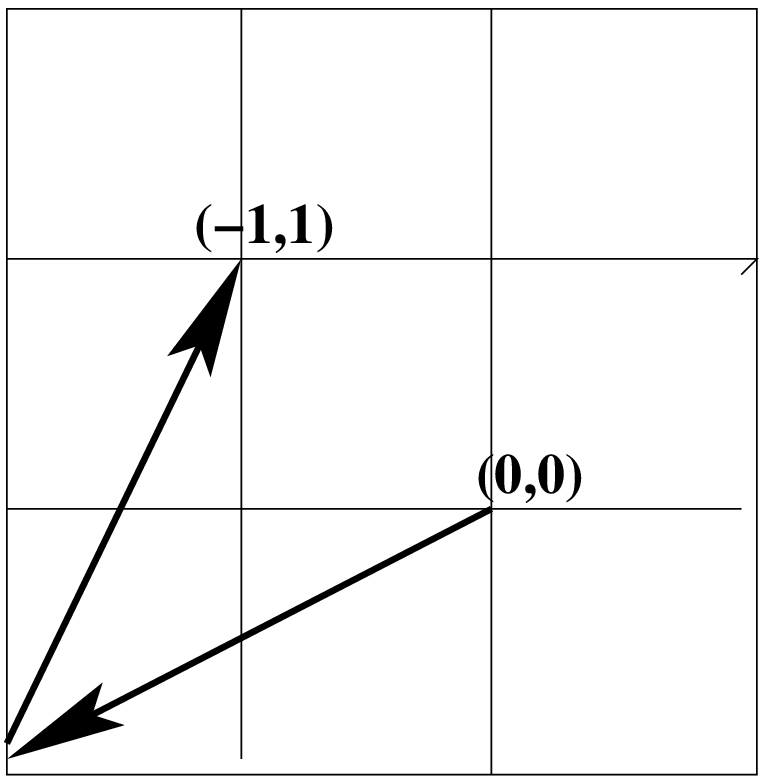}
\vspace{.5cm}
\noindent (b)~$Q=(\frac{2}{3},\frac{1}{3}), p_1 = (1,2), p_2 = (2,1)$
\hfill
\includegraphics[width=2.5cm]{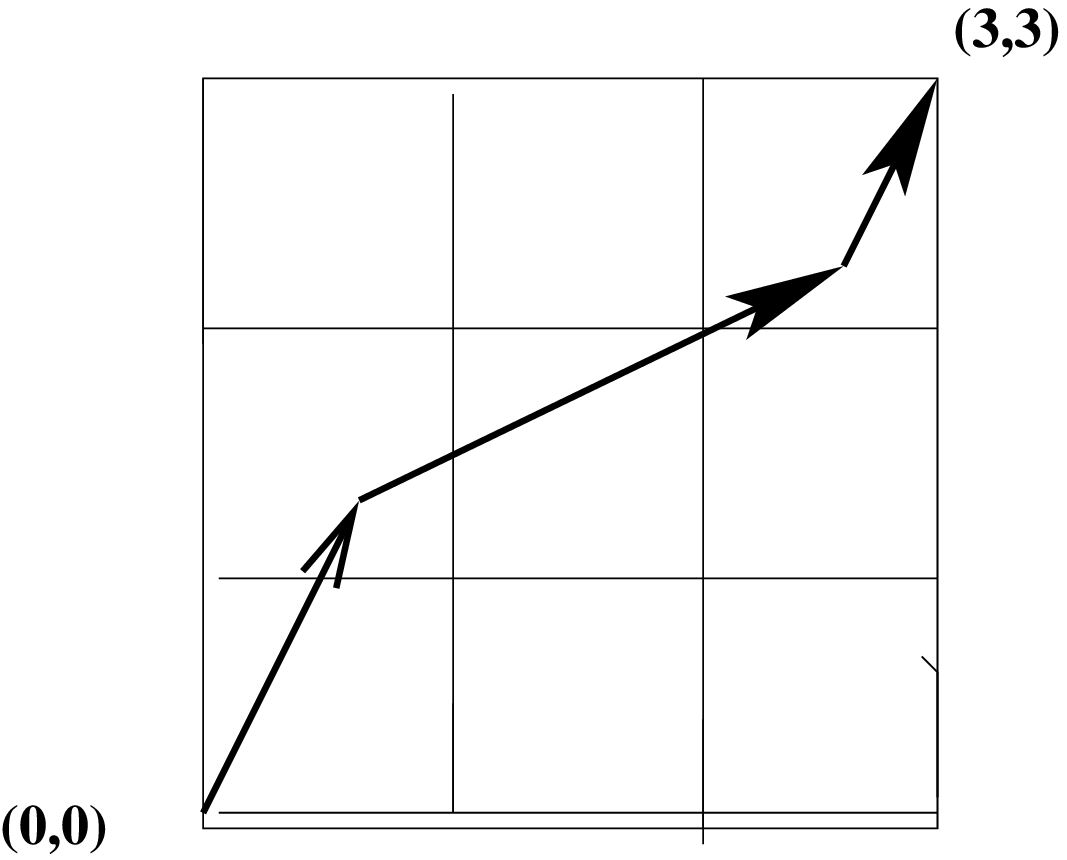}
\hspace{0.5cm}
\includegraphics[width=2cm]{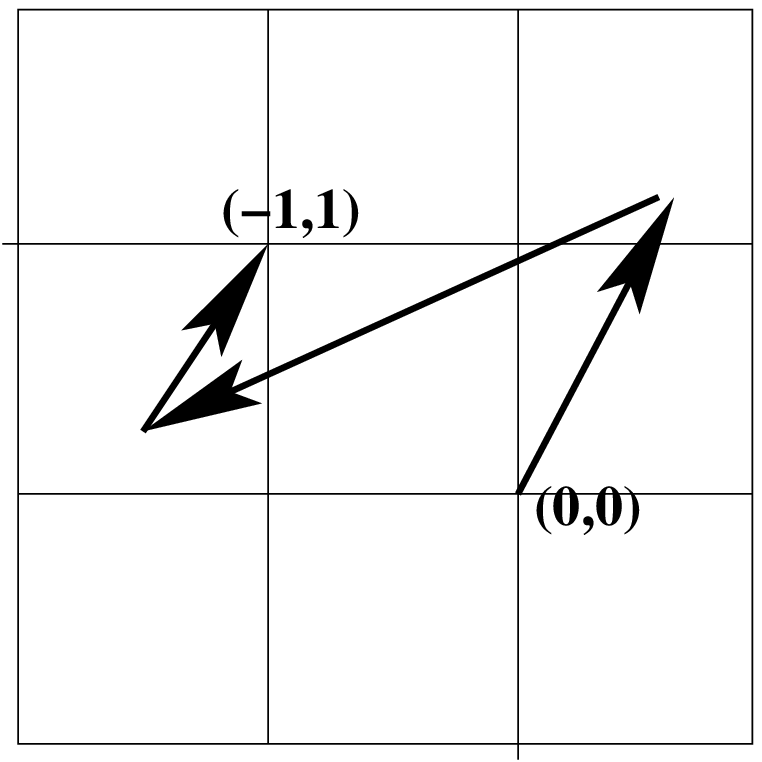}
\vspace{.5cm}
\noindent (c)~$R=(\frac{1}{3},\frac{2}{3}), p_1 = (1,2), p_2 = (2,1)$
\hfill
\includegraphics[width=2.5cm]{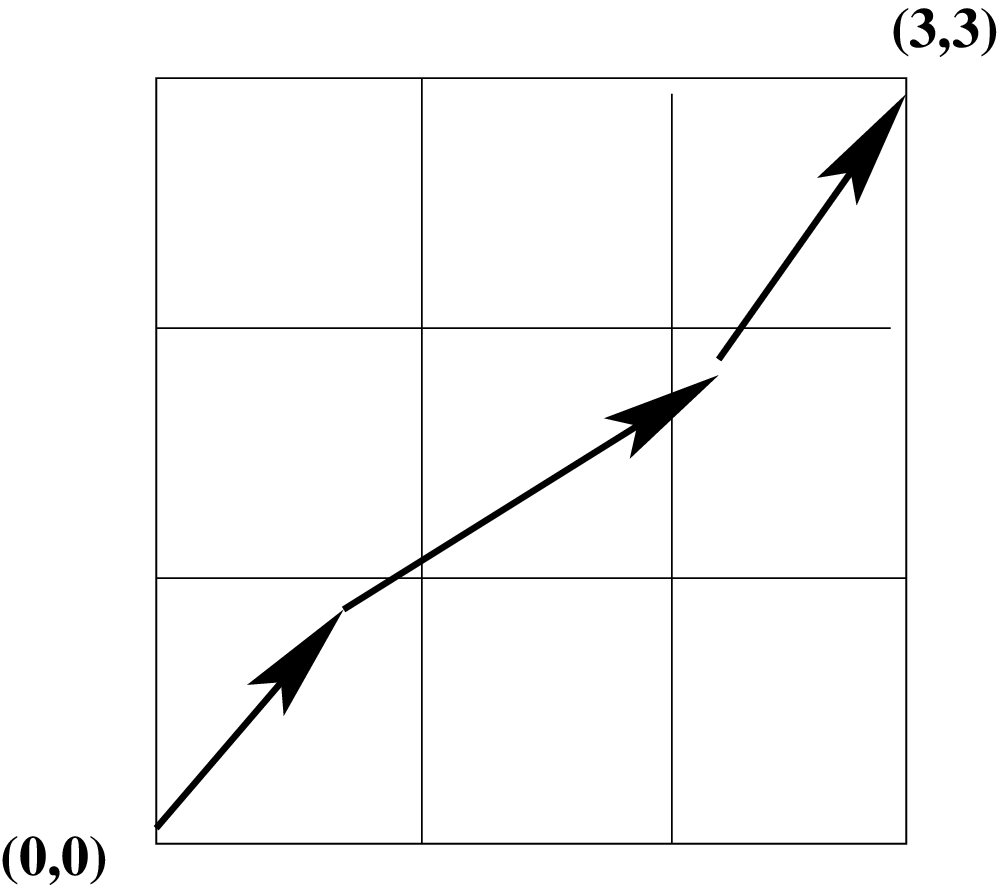}
\hspace{0.5cm}
\includegraphics[width=2cm]{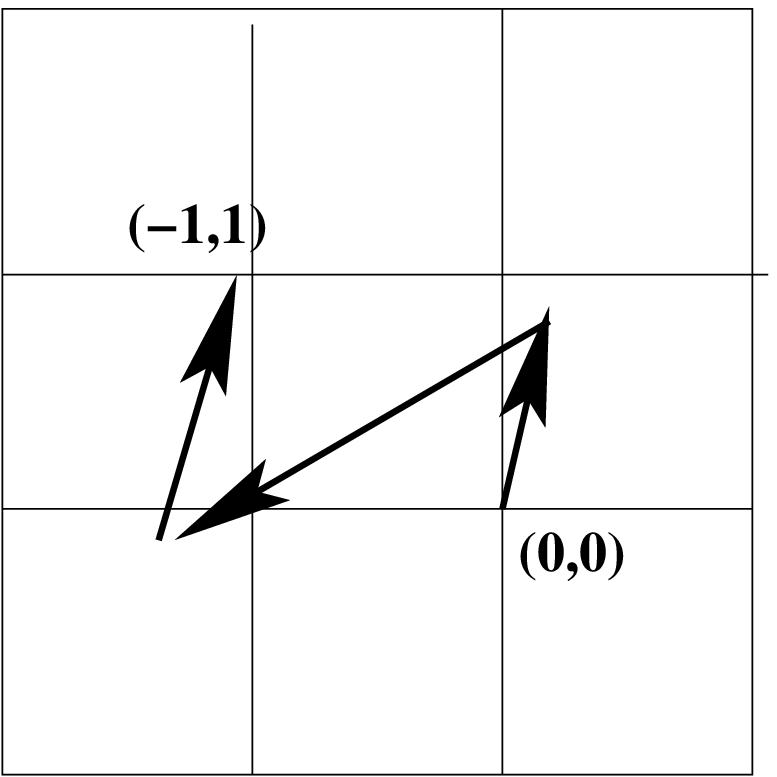}

\item (a)~$ P=(0,0), p_1 = (1,1), p_2 =(-1,2)$
\hfill
\includegraphics[width=1.5cm]{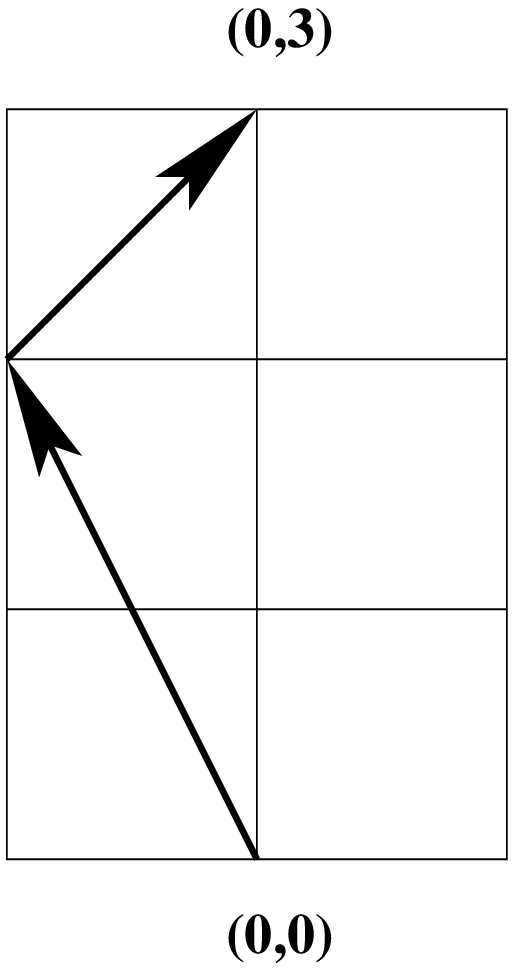}
\hspace{0.5cm}
\includegraphics[width=2.7cm]{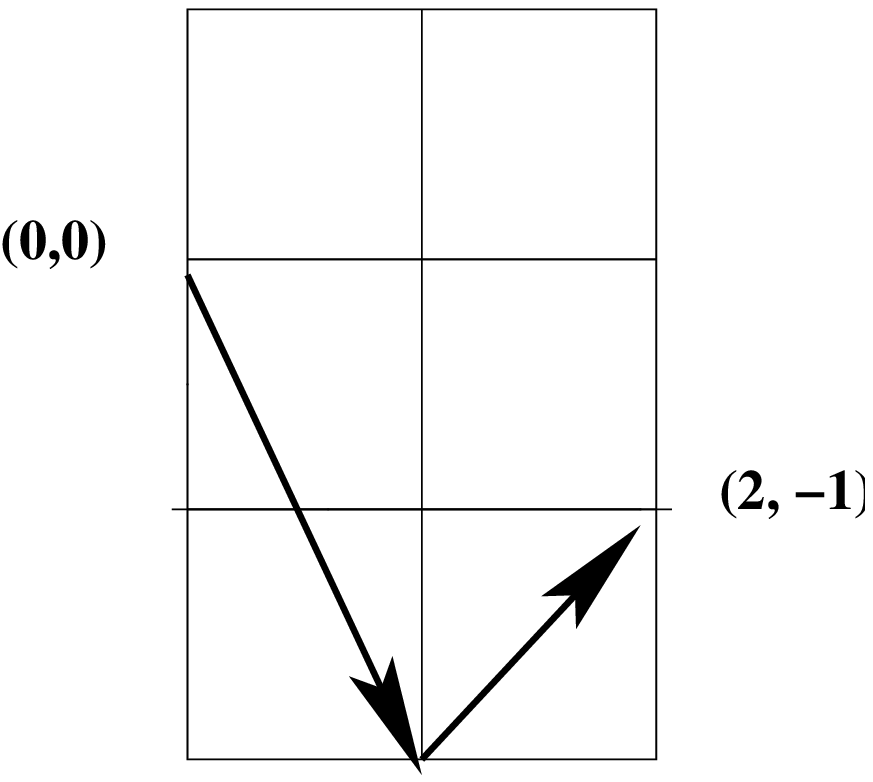}
\vspace{.5cm}
\noindent (b)~$Q=(\frac{2}{3},\frac{2}{3}), p_1 = (1,1), p_2 = (-1,2)$
\hfill
\includegraphics[width=1.5cm]{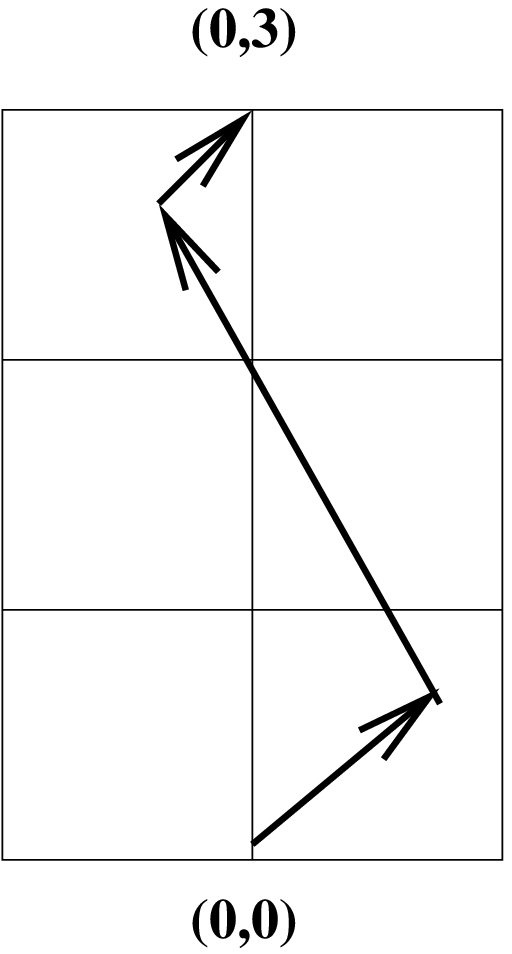}
\hspace{0.5cm}
\includegraphics[width=2.7cm]{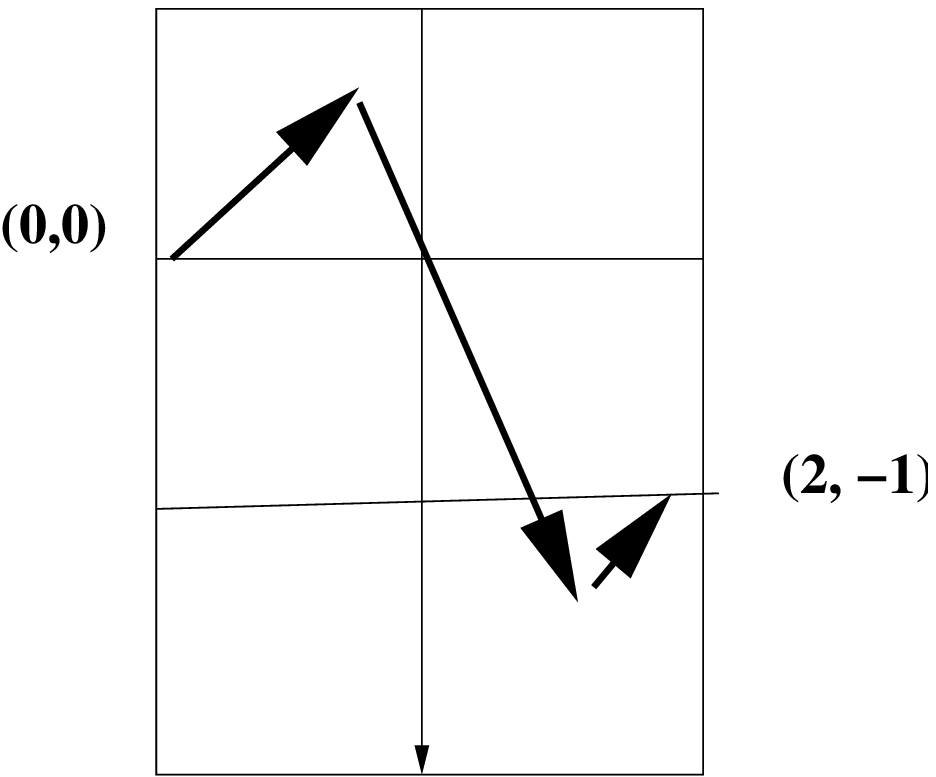}
\vspace{.5cm}
\noindent (c)~$R= (\frac{1}{3},\frac{1}{3}), p_1 = (1,1), p_2 = (-1,2)$
\hfill
\includegraphics[width=1.5cm]{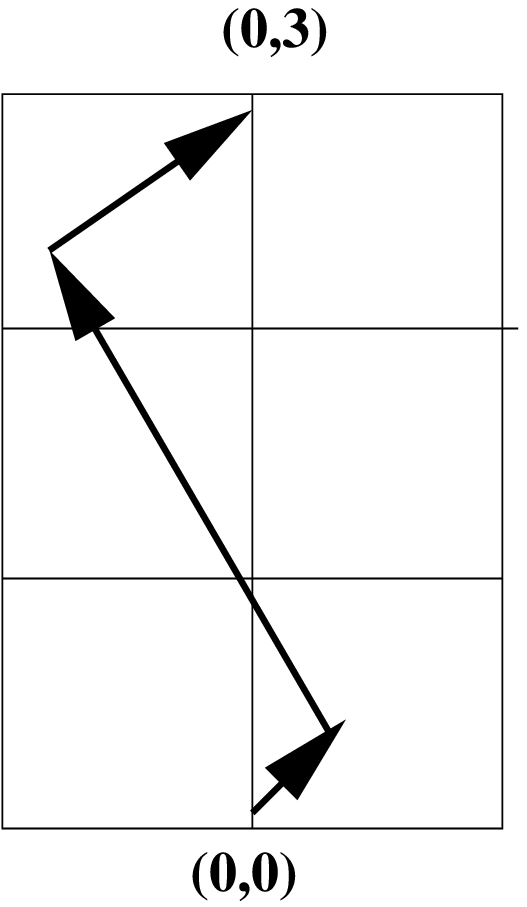}
\hspace{0.5cm}
\includegraphics[width=2.7cm]{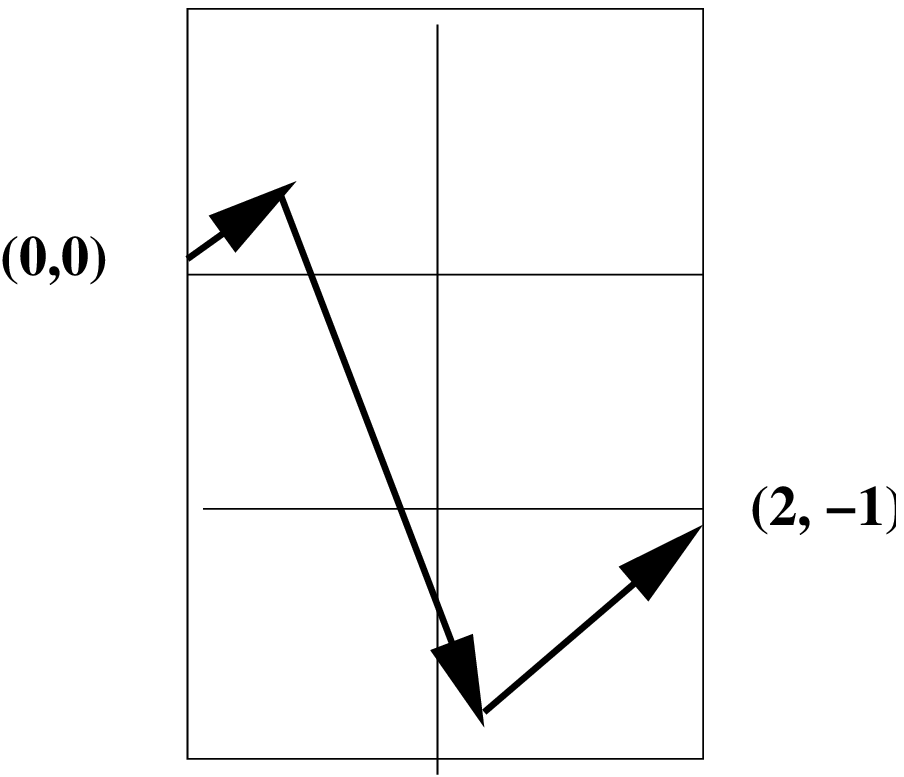}
\end{enumerate}

In each of the above examples, and for each intersection point $P, Q, R$, it
is clear that $p_1Pp_2\sim (m+s,n+t)$ and $p_1Pp_2^{-1}\sim(m-s,n-t)$.

\subsection{The Goldman bracket\label{goldbrac}}
We can assign classical functions to the straight paths $(m,n)$
of Section \ref{fundred} as follows
\be
T(m,n) = e^{mr_1 + nr_2} + e^{- mr_1 - nr_2},
\label{tch}
\ee
i.e. $T(m,n) = {\rm tr}~~ U_{(m,n)}$ where $U_{(m,n)}$ is of the form
\rref{Umn} with $r_1,\,r_2$ classical parameters. In terms of the
Poisson bracket \rref{pb} the Poisson bracket between these functions 
for two paths $(m,n)$ and $(s,t)$ is
\be
\{T(m,n), T(s,t)\} = (mt-ns)(T(m+s,n+t) - T(m-s,n-t))\{r_1,r_2\}
\label{pbt}
\ee

Equation (\ref{pbt}) may be regarded as a particular case of the Goldman
bracket \rref{gold} (modulo rescaling $\{r_1,r_2\}$ to $1$),  since $(m,n)$
and $(s,t)$ have total intersection index $mt-ns$, and the rerouted paths $p_1Qp_2$
and $p_1Qp_2^{-1}$, where $p_1=(m,n)$ and $p_2=(s,t)$, are all
homotopic to $(m+s,n+t)$ and $(m-s,n-t)$ respectively.

The bracket \rref{pbt} can be quantized using the triangle identity (see \rref{tri})
\be
e^{mr_1 + nr_2} e^{sr_1 + tr_2} = q^{(mt-ns)/2} e^{(m+s)r_1 + (n+t)r_2}
\ee
and the result is the commutator
\be
[ T(m,n), T(s,t)] = (q^{\frac{(mt-ns)}{2}} - q^{-\frac{(mt-ns)}{2}})
(T(m+s,n+t) - T(m-s,n-t)).
\label{tcomm} 
\ee
The antisymmetry of \rref{tcomm} is evident (from \rref{tch} 
$T(m,n)=T(-m,-n)$). It can be checked that \rref{tcomm} satisfies the Jacobi 
identity, and that the classical limit $q \to 1, \hbar \to 0$
$$
\{,\} = lim_{\hbar \to 0}\frac{[,]}{i \hbar}
$$
of (\ref{tcomm}) is precisely \rref{pbt}.

We will now give a different version of equation \rref{tcomm} which treats 
each intersection point individually, and uses rerouted paths homotopic to 
``straight line'' paths as discussed in Section \ref{rerout}, since we have 
already seen in Section \ref{sec3} that homotopic paths no longer have the
same  quantum matrix assigned to them, but only the same  matrix up to a
phase. Thus for an arbitrary PL  path $p$ from
$(0,0)$ to $(m,n)$, set
\be
T(p) = q^{S(p,(m,n))} T(m,n).
\label{tfactor}
\ee
The factor appearing in \rref{tfactor} is the same as that relating the 
quantum matrices $U_p$ and $U_{(m,n)}$, where $(m,n)$ is the straight path. 

\begin{figure}[hbpt]
\centering
\includegraphics[height=3cm]{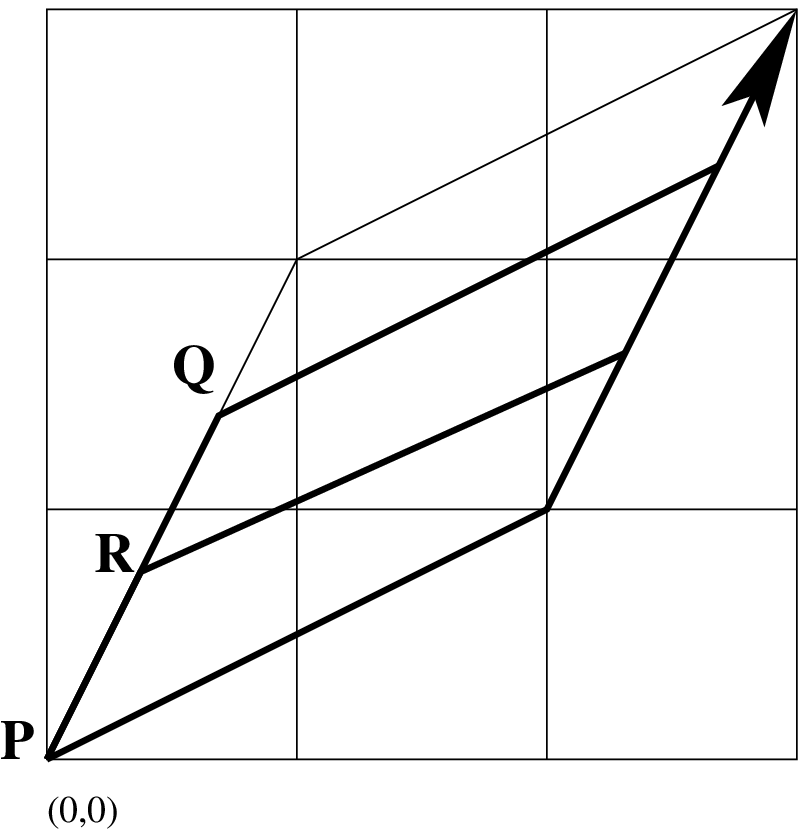}
\hspace{2cm}
\includegraphics[height=3cm]{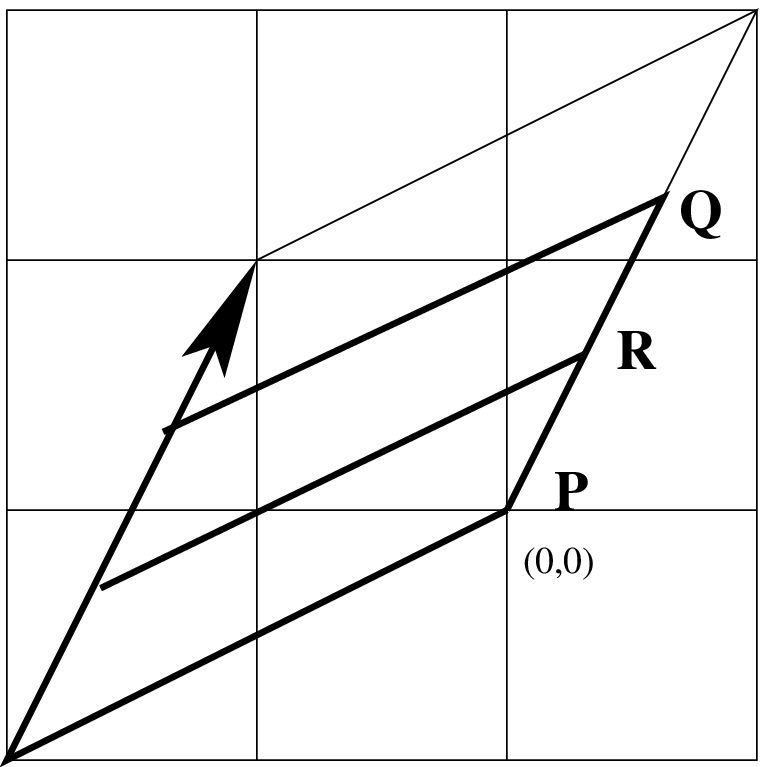}
\caption{ The reroutings $(1,2)S(2,1)$ and $(1,2)S(-2,-1)$ for $S=P,R,Q$ }
\label{p20a}
\end{figure}

Firstly we will show how to rewrite \rref{tcomm} in terms of the rerouted
paths of Section \ref{rerout} for one of the previous examples (example 3 of
Section \ref{ints}). In this example  $p_1=(1,2),\,p_2=(2,1)$, and we
have, from \rref{tcomm}
\be
[ T(1,2), T(2,1)] = (q^{-3/2} - q^{3/2})(T(3,3) - T(-1,1)).
\label{tcommexp} 
\ee 
The intersections occur at the points $P, R, Q$ (in that order,
counting along $p_1$) as shown in Figure \ref{p20a}. For the positively
rerouted paths at these points we have the following equations
\bea
T((1,2)P(2,1)) & = & T((2,1)(1,2))=q^{3/2} T(3,3) \label{P}\\
T((1,2)R(2,1))& = & q^{-1}T((1,2)P(2,1)) \label{R}\\
T((1,2)Q(2,1)) & = & q^{-1} T((1,2)R(2,1)) \label{Q}
\end{eqnarray}
and for the negative reroutings
\bea
T((1,2)P(-2,-1)) & = & T((-2,-1)(1,2))=q^{-3/2} T(-1,1) \label{P-}\\
T((1,2)R(-2,-1))& = & q T((1,2)P(-2,-1)) \label{R-}\\
T((1,2)Q(-2,-1)) & = & q T((1,2)R(-2,-1)) \label{Q-}.
\end{eqnarray}

\begin{figure}[hbtp]
\centering
\includegraphics[height=3cm]{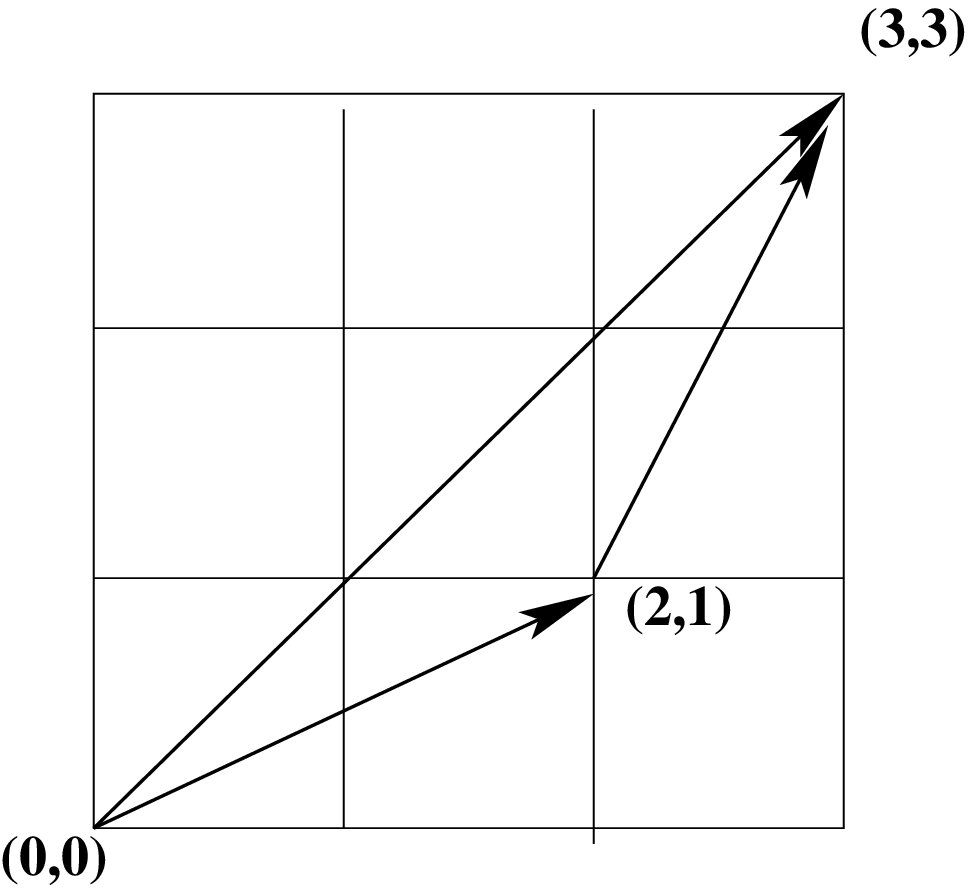}
\hspace{2cm}
\includegraphics[height=3cm]{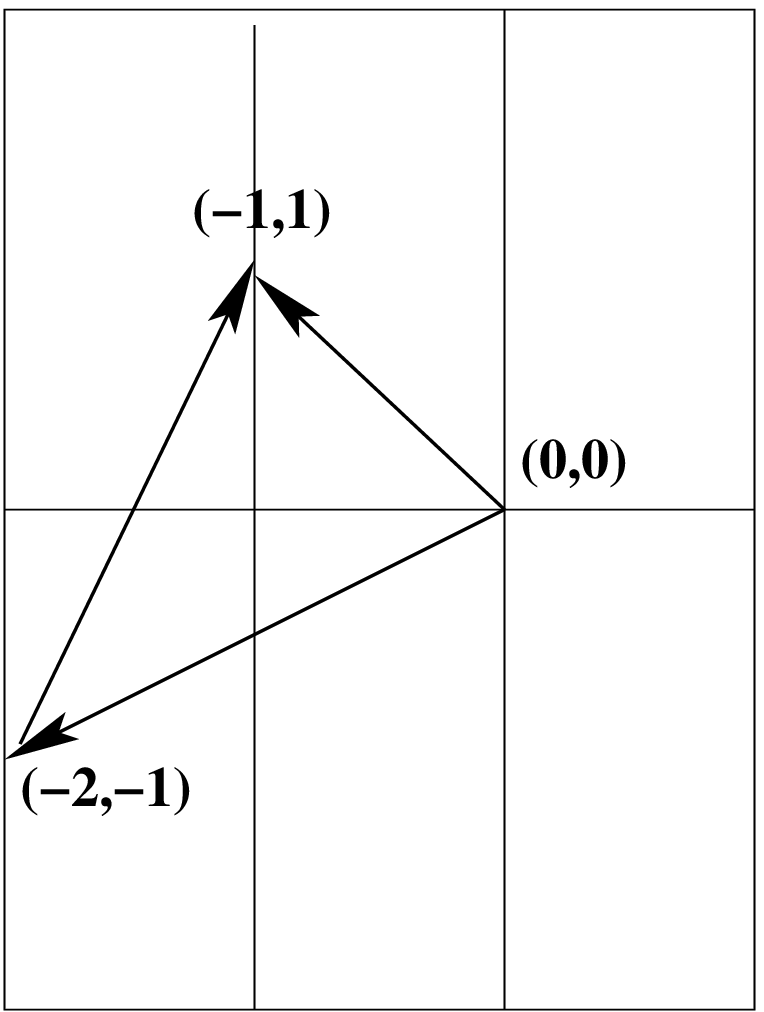}
\caption{Factors for the reroutings $(1,2)P(2,1)$ and $(1,2)P(-2,-1)$}
\label{trigs}
\end{figure}

The factors appearing in equations \rref{R}, \rref{Q}, \rref{R-} and \rref{Q-}
(the reroutings at the points $R$ and $Q$) are shown in Figure \ref{p20a}, 
where it is clear that each large parallelogram is divided into
three equal parallelograms, each of unit area. The factors  in \rref{P} and
\rref{P-} (the reroutings at the point $P$) come from the  triangle equation
\rref{tri}, and are shown in Figure \ref{trigs}, where the triangles have
signed area $+\frac{3}{2}$ and $-\frac{3}{2}$ respectively.

After a simple calculation we find that equation \rref{tcommexp} can be 
rewritten in the form:
\be
[T(1,2),T(2,1)] =\sum_{S=P,R,Q} 
(q^{-1} -1) T((1,2)S(2,1)) + (q-1) T((1,2)S(-2,-1)). 
\label{qgoldexp}
\ee

In the general case, for $p_1=(m,n)$ and $p_2=(s,t)$ with $mt-ns\neq 0$, 
we postulate that
\be
[T(p_1), T(p_2)] = \sum_{ Q \in p_1 \sharp p_2}
(q^{\epsilon(p_1,p_2,Q)} - 1)T(p_1Qp_2)  
+ (q^{-\epsilon(p_1,p_2,Q)} - 1)T(p_1Qp_2^{-1})
\label{qgold}
\ee
quantizes the Goldman bracket \rref{gold}.

\begin{figure}[hbpt]
\centering
\includegraphics[height=4cm]{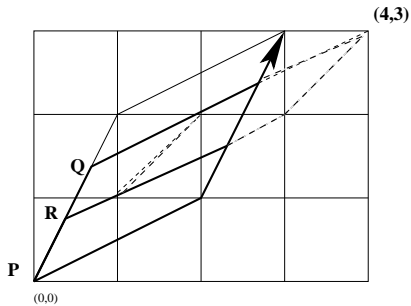}
\caption{ The area of the middle parallelogram is $1$}
\label{pp20a}
\end{figure}

To prove equation \rref{qgold} we first assume that both $p_1$ and $p_2$
are irreducible, i.e. not multiples of other integer paths, 
and study the reroutings $p_1 Qp_2$ with $Q$ an intersection
point, as discussed in Section \ref{rerout}. They are paths similar to those
of Figure \ref{p20a}, namely  following $p_1$ to $Q$, then rerouting along a
path {\it parallel} to $p_2$, then finishing along a path {\it parallel} to
$p_1$. The reroutings along $p_2$ obviously must pass through an integer
point inside the parallelogram formed by $p_1$ and $p_2$ (apart from when
the intersection point is the origin). They also clearly pass through only one
integer point since $p_2$ is irreducible. Consider two adjacent lines inside
the parallelogram parallel to $p_2$ and passing through integer points. 
We claim that the area of each parallelogram between them is $1$.
Consider for instance one of the middle parallelograms in 
Figure \ref{p20a} (whose area we saw previously was $1$ as the three 
parallelograms are clearly of equal area and the area of the large 
parallelogram is $3$). 
This is the same area as that of a parallelogram with vertices at
integer points, as can be shown, for example, by cutting it into two pieces 
along the line between (1,1) and (2,2), then regluing them together into a 
parallelogram with vertices at (1,1), (2,2), (3,2) and (4,3), as indicated in
Figure \ref{pp20a}. This latter area is equal to $1$ from Pick's theorem
\cite{pick} which states that the area $A(P)$ of a lattice polygon $P$ is
\be A(P) = I(P) + B(P)/2 - 1, \label{pick} 
\ee 
where $I(P)$ is the number of interior lattice points and $B(P)$ is the number
of boundary points 
(for the parallelogram in the example $I(P)=0$ since
the lines parallel to $p_2$ are adjacent, and $B(P)=4$ from the integer
points at the four vertices, so $A(P) = 0+4/2-1=1$.) Therefore in general the
parallelogram determined by $p_1$ and $p_2$, whose total area is $A=
|mt-ns|$, is divided up into $A$ smaller parallelograms of equal area by lines
parallel to $p_2$ passing through the interior integer points of the
parallelogram. The fact that the total area is equal to the number of internal
integer points $+1$ is again a consequence of Pick's theorem. 

We can now calculate the first term (the positive reroutings shown for the
example in Figure \ref{p20a}) in the sum on the r.h.s. of
\rref{qgold},  using equation \rref{tfactor}, and show that it is equal to the
first term on the  r.h.s. of \rref{tcomm}. Consider first the case
$\epsilon(p_1,p_2,Q) = -1$. The  rerouting at the origin satisfies, using
the triangle equation \rref{tri}, 
\[
T(p_1\, (0,0)\, p_2) = q^{A/2} T(m+s,n+t),
\]
where the area of the parallelogram determined by $p_1,\,p_2$ is $A=-(mt-ns)$.  
The next rerouted path adjacent to $p_1\, (0,0)\, p_2$, rerouted at $Q_1$ say 
(in the example $Q_1=R$) satisfies 
\[
T(p_1\, Q_1\, p_2) = q^{-1}T(p_1\, (0,0)\, p_2)
\]
since we have shown that the signed area between the paths is $-1$. Similarly 
each successive adjacent path rerouted at $Q_2, Q_3, \dots$ satisfies
\be
T(p_1\, Q_i\, p_2) = q^{-1}T(p_1\, Q_{i-1}\, p_2).
\ee
with $Q_0$ the origin $(0,0)$. It follows that 
\begin{eqnarray}
\lefteqn{\sum_{Q \in p_1 \sharp p_2}(q^{-1}-1)T(p_1 Q p_2)}\nonumber \\ 
&=& (q^{-1}-1)q^{A/2}(1+q^{-1} + \dots + q^{-(A-1)})T(m+s,n+t) \nonumber \\
&=& (q^{-1}-1)q^{A/2}\frac{1-q^{-A}}{1-q^{-1}}T(m+s,n+t)\nonumber \\
&=& (q^{-A/2} - q^{A/2})T(m+s,n+t)\nonumber \\
&=& (q^{(mt-ns)/2} - q^{-(mt-ns)/2})T(m+s,n+t).
\label{ep+}
\end{eqnarray}
When $\epsilon(p_1,p_2,Q)=+1$ the calculation is identical to \rref{ep+} 
but with $q$ rather than $q^{-1}$, 
and with the area of the triangle now equal to $A/2$, where $A=mt-ns$, 
namely
\begin{eqnarray}
\lefteqn{\sum_{ Q \in p_1 \sharp p_2}(q-1) T(p_1\, Q\, p_2)} \nonumber \\
&=& (q-1)q^{-A/2}(1+q^{1} + \dots + q^{A-1}) T(m+s,n+t)\nonumber \\
&=& (q^{A/2} - q^{-A/2}) T(m+s,n+t)\nonumber \\
&=& (q^{(mt-ns)/2} - q^{-(mt-ns)/2}) T(m+s,n+t).
\label{ep-}
\end{eqnarray}

Diagrammatically this corresponds to dividing up the first parallelogram in Figure
\ref{p20a} by lines passing through the integer points in the interior, but
parallel to $(1,2)$, as opposed to $(2,1)$.

In an entirely analogous way the second terms (the negative reroutings) on the 
r.h.s. of \rref{tcomm} and \rref{qgold} can be shown to be equal - the second
figure of Figure \ref{p20a} can be used as a guide\footnote {The antisymmetry
of \rref{qgold} can be checked for our example $p_1=(1,2),p_2=(2,1)$, both
irreducible, by noting that the intersections  occur at the same points (but in
a different order, namely  $P, Q, R$).}.

When $p_1$ is reducible, i.e. $p_1 =c(m',n'), \, c\in \mathbb{N}, m', n' \in
\mathbb{Z}$, and $p_2$ is irreducible, formula \rref{qgold} applies exactly as
for the irreducible case, since there are $c$ times as many rerouted paths
compared to the case when $p_1=(m',n')$. An example is $p_1=(2,0),\,
p_2=(1,2)$, where the first term on the r.h.s. of \rref{tcomm} is equal to the 
first term on the r.h.s. of \rref{qgold}:
\begin{eqnarray}
(q^2-q^{-2}) T(3,2) & = & (q-1)q^{-2} (1+q+q^2 + q^3) T(3,2)\nonumber \\
&=& \sum_{ Q \in p_1 \sharp p_2} (q-1) T(p_1Q p_2).
\end{eqnarray}
There are four rerouted paths in the final summation rerouting at $(0,0)$,
$(1/2,0)$, $(1,0)$ and  $(3/2,0)$ along $p_1$.

If $p_2$ is reducible we must use multiple intersection numbers
in \rref{qgold},
 i.e. not simply $\pm 1$. Suppose $p_1=(m,n)$ and $p_2=(s,t) = c(s',t')$, $
c\in \mathbb{N}, s', t' \in \mathbb{Z}$. Then for example the first
term on the r.h.s. of \rref{qgold} with $mt-ns>0$ is 
\begin{eqnarray}
\lefteqn{(q^{(mt-ns)/2}) - q^{-(mt-ns)/2}) T(m+s,n+t)} \nonumber \\
&=& (q^{c(mt'-ns')}- 1) q^{-(mt-ns)/2} T(m+s,n+t) \nonumber\\
&=& (q^c-1) q^{-(mt-ns)/2} (1 + q^c + \dots + q^{c(mt'-ns'-1)}) T(m+s,n+t)\nonumber \\
&=& \sum_{ Q \in p_1 \sharp p_2} (q^c-1)  T(p_1Q p_2).
\label{multint}
\end{eqnarray}

The factor $(q^c-1)$ is the {\it quantum} multiple intersection number at the
$mt'-ns'$ intersection points. 
The calculation can be regarded as doing equation \rref{ep-} backwards
and substituting $q$ by $q^c$ and $mt-ns$ by $mt'-ns'$.
 An example is given by $p_1=(2,1),\, p_2
= (0,2)$, for which double intersections occur along $p_1$ at the origin
and at $(1,1/2)$. From \rref{multint} the first term on the r.h.s. of
\rref{qgold} is

\begin{eqnarray}
(q^2-q^{-2}) T(2,3) & = & (q^2-1)q^{-2} (1+q^2) T(2,3) \nonumber \\
&=& (q^2-1) (T(p_1\, (0,0)\,Q p_2) + T(p_1\, (1,1/2)\, p_2).
\end{eqnarray}

\section{Conclusions\label{sec7}}

The quantum geometry emerging from the use of a constant quantum connection
exhibits some surprising features. The phase factor appearing in the
fundamental relation \rref{fundsl2} acquires a geometrical origin as the signed
area phase relating two integer PL paths, corresponding to two different loops 
on the torus e.g. as in Figure \ref{fundexp}. The signed area
phases between PL paths have good properties, and lead to a natural concept of
$q$-deformed surface group representations. The classical correspondence
between flat connections (local geometry) and holonomies, i.e. group
homomorphisms from $\pi_1$ to $G$ (non-local geometry) thus has a natural
quantum counterpart. It is interesting to speculate that the signed area phases
could be related to gerbe parallel transport \cite{MP}.  

The signed area phases also appear in a quantum version \rref{qgold} of a
classical bracket \rref{gold} due to Goldman \cite{gol}, where
classical  intersection numbers $\pm \ep(p_1,p_2,Q)$ are replaced by quantum
single and multiple intersection numbers $(q^{\pm \ep(p_1,p_2,Q)}-1)$. 

The quantum bracket for homotopy classes represented by straight lines
\rref{tcomm} is easily checked since all the reroutings are homotopic. However
the r.h.s. of the bracket \rref{qgold} is expressed in terms of rerouted paths
using the signed area phases and a far subtler picture emerges. 

The quantum bracket for straight lines \rref{tcomm} is manifestly
antisymmetric. Although we have shown that \rref{tcomm} and \rref{qgold} are
equivalent the antisymmetry of \rref{qgold} is not at all obvious. This
antisymmetry can be checked however when both paths are irreducible e.g. the
example in Section \ref{goldbrac}, even though the intersections in the two
cases may occur in a different order. When one of the paths is reducible then
the r.h.s. is expressed in terms of either multiple reroutings with single
intersection numbers (for $p_1$ reducible) or single reroutings with multiple
intersection  numbers (for $p_2$ reducible). The case when both $p_1$ and $p_2$
are reducible is a combination of the above. There should be an interplay
between the two scenarios.

It is not difficult to show that the Jacobi identity holds for the quantum 
bracket for straight paths \rref{tcomm} since the r.h.s. may also be expressed
in terms of straight paths, with suitable phases. It must also hold for
\rref{qgold} since they are equivalent. To show this explicitly for arbitrary
PL paths implies extending the bracket \rref{qgold} to arbitrary PL paths,
without identifying homotopic paths. It should also be possible to treat higher
genus surfaces (of genus $g$) in a similar fashion by introducing the same
constant quantum connection on a domain in the $xy$ plane bounded by a $4g$--gon
with the edges suitably identified \cite{hil}. One could then define holonomies
of PL loops on this domain and study their behaviour under intersections, in an
analogous fashion to the genus $1$ case studied here. These, and related
questions, will be studied elsewhere.

\bibliographystyle{my-h-elsevier}

\end{document}